\documentclass[a4paper,11pt]{article}

\usepackage{booktabs} 

\usepackage{fullpage}
\usepackage[T1]{fontenc}
\usepackage{comment}

\usepackage{color}
\usepackage{graphicx}
\usepackage{epsf}
\usepackage{graphicx,epsfig}

\usepackage{multirow}

\usepackage{amsmath,amsthm,amssymb}

\usepackage{physics}
\usepackage{hyperref}
\usepackage{prettyref}

\usepackage{epsfig}
\usepackage{cite}
\usepackage{color,colordvi}
\usepackage{xcolor}

\newcommand{\be}{\begin{eqnarray}}
\newcommand{\ee}{\end{eqnarray}}

\numberwithin{equation}{section}

\begin{document}

\setcounter{footnote}{0}

\baselineskip 6 mm

\begin{titlepage}
	\thispagestyle{empty}
	\begin{flushright}
		
	\end{flushright}

	\vspace{35pt}
	
	\begin{center}
	    { \Large{\bf G2 resolutions and backtracking in scale-separated AdS$_3$}} 
		
		\vspace{50pt}
		
		{Aviral Aggarwal$^1$, Fien Apers$^2$, Fotis Farakos$^3$}  
		
		\vspace{25pt}

     $^1${\it Department of Theoretical Physics and Astrophysics, \\ 
    Faculty of Science, Masaryk University, 611 37 Brno, Czech Republic.}
		
		\vspace{15pt}

    $^2${\it Instituto de F\'isica Te\'orica IFT-UAM/CSIC, C/ Nicol\'as Cabrera 13-15, \\ 
    Campus de Cantoblanco, 28049 Madrid, Spain}
		
		\vspace{15pt}

    $^3${\it Physics Division, National Technical University of Athens \\
        15780 Zografou Campus, Athens, Greece}
		
		\vspace{15pt}

		\vspace{40pt}
		
		{ABSTRACT} 
	\end{center}

We study scale-separated AdS$_3$ models with full moduli stabilization in massive IIA supergravity on G2 orientifolds. In addition to constructing new examples, we investigate the properties of these models once the singularities of the G2 toroidal orbifold are resolved. The blow-up moduli are shown to be stabilized using fluxes in a parametric regime that preserves scale separation and control under the supergravity approximation. We further show that the apparent orientifold intersections in the singular orbifold description do not exist on such resolved spaces. Finally, we analyze the brane dual picture of such models using flux backtracking. We highlight several interesting features from such analysis such as a finite interval solution of the resulting flow equations in some of the models and the desirable low-energy decoupling of brane modes.

\vspace{10pt}

\bigskip

\end{titlepage}

\newpage

\tableofcontents

\newpage

\section{Introduction}

Flux compactifications of string theory/M-theory on spaces with G2 structure \cite{Joyce:2002eb,Joyce1,Joyce2} have received a lot of attention as a playground for constructing phenomenologically relevant models 
\cite{Acharya:1996fx,Acharya:1998pm,Atiyah:2001qf,Friedmann:2002ty,Cvetic:2001nr}. 
Recently, owing to their relatively simple description in terms of toroidal orbifolds, 
minimal supersymmetry preservation, and Ricci flatness, 
G2 orientifolds have also been studied as potential candidates for classical supersymmetric AdS vacua with parametric scale separation \cite{Coudarchet:2023mfs}. 
These studies cover a wide range of the string theory/M-theory web: 
constructions with G2 holonomy in massive IIA \cite{Farakos:2020phe,VanHemelryck:2022ynr,Aikot:2026tzn} which yield $\mathcal{N}=1$ AdS$_3$, 
constructions with torsion in massless IIA lifted to M-theory \cite{Cribiori:2021djm,VanHemelryck:2024bas,Tringas:2026ncg} which yield $\mathcal{N}=1$ AdS$_4$ or AdS$_3$, and
constructions in IIB \cite{Arboleya:2025lwu,Arboleya:2024vnp,VanHemelryck:2025qok,Arboleya:2025jko} which lead to scale-separated AdS$_3$, 
while there are more recent equivalent results in heterotic string theory \cite{Tringas:2025bwe} and in type I \cite{Miao:2025rgf}.

At this stage, there exist several preliminary results that show that the low-energy effective supergravity description of such compactifications is a valid approximation and not simply a consistent truncation.
The various constructions make use of the typical zeroth-order approximations and assumptions, while the details are unavoidably left for further checks.
Recent developments aim to understand these approximations and work out the details so that scale separation is put on firmer ground.
For example, one notable approximation is the smeared backreaction of orientifold sources.
However, this has now been studied to leading order in the localized backreaction, 
supporting that the constructions hold water \cite{Junghans:2020acz,Marchesano:2020qvg,Emelin:2022cac}, 
while some further considerations for the restrictions from higher orders have been studied in \cite{Emelin:2024vug}. 
Another typical feature was the presence of Romans mass in the original DGKT-CFI constructions \cite{DeWolfe:2005uu,Camara:2005dc}; 
however, more recent constructions have bypassed this using torsion \cite{Cribiori:2021djm}. 
Moreover, it has been shown in \cite{Junghans:2023yue} that orientifold intersection issues are eliminated by carefully examining the properties of the internal space.
Indeed, in the resolved internal space, no intersecting orientifolds appear, in contrast to the orbifold limit \cite{DeWolfe:2005uu,Camara:2005dc}. 
A series of further checks have been performed, ranging from backtracking solutions \cite{Apers:2022vfp,Apers:2025pon,Apers:2026ahu,Arboleya:2026irl,Apers:2026lgi} to consistency checks based on constraints imposed by the putative dual conformal field theory \cite{Bedroya:2025ltj,Bobev:2025yxp,Revello:2026eqp}.

In this work, our aim is to address a few more open questions that have been left unresolved in previous G2 constructions.
Our main focus is on building a model with G2 holonomy and a tractable moduli spectrum so that we can analyze full moduli stabilization, including the blow-up modes.
To this end, we work with G2 orbifolds with known resolutions within the Joyce family \cite{Joyce2}, and require full moduli stabilization in the untwisted sector before we work out the blow-ups (which typically belong to the twisted sector).
This has the advantage that blow-up moduli stabilization can be treated as a perturbation on top of the already stabilized untwisted sector.
This is a non-trivial feature, but we have identified a specific orbifold that provides a working example and have also checked a variety of flux choices that lead to full moduli stabilization (all this is done in Section~\ref{Section 2:G2 orbifolds and moduli stabilization} for our main flux example, 
while other flux choices are studied in the appendix).

Once we have a working model, we proceed with the analysis of the twisted sector that arises from the orbifold desingularization. 
The main questions we want to address are how the blow-up moduli get stabilized (Section~\ref{Section 3: Twisted sector moduli stabilization}), 
and to further check that, once the internal space is smooth, the orientifolds cease to intersect (Section~\ref{Section 4 :Impact of the resolution}). 
We also analyze some non-trivial features of tadpole cancellation related to the resolution of the orbifold singularities. 

In Section~\ref{Section 5:The brane dual}, we study the brane dual of our working example. Using flux backtracking \cite{Apers:2025pon}, we derive an analytic expression for the singular geometry to be probed by a network of D4- and D2-branes supporting the putative CFT dual on their worldvolume. The geometry takes the form of a \emph{finite interval} bounded by two singular endpoints, with the branes located at the strongly coupled endpoint. As in the DGKT AdS$_4$ vacua, these branes decouple from the bulk in the low-energy limit, satisfying a necessary condition for a holographic description \cite{Apers:2026lgi,Apers:2026ahu}. The finite extent of the interval also avoids the potential asymptotic pathologies encountered in other known scale-separated constructions \cite{Bedroya:2025ltj}. Although the complexity of the D4/D2-brane network, including Freed--Witten effects, prevents us from obtaining an analytic solution for the full brane geometry, we construct a numerical solution interpolating between the AdS$_3$ minimum and the weakly coupled endpoint.

Finally, we present some further technical details and more examples in the appendices.

\section{G2 orbifolds and moduli stabilization}\label{Section 2:G2 orbifolds and moduli stabilization}

Here we review the G2 orbifolds presented in \cite{Farakos:2020phe,Aikot:2026tzn}, discussing a G2 orbifold that has not previously been considered in studies of scale separation in massive IIA.
These provide a favorable playground to explore flux compactifications down to 3D that preserve supersymmetry.
Among all the different possible realizations of G2 manifolds, our choice is simply dictated by the fact that it provides a good working example.

\subsection{Standard contributions}

The basis of our analysis will be the G2-orientifold constructions presented in \cite{Farakos:2020phe,Aikot:2026tzn}, which we will rapidly review here, also taking the opportunity to clarify conventions and notation.
We will be working with massive IIA supergravity with O2/O6 planes and possibly D-branes.  
A self-consistent setup can contain $F_0$ and $F_4$ RR fluxes, as well as non-zero $H_3$ flux in the internal space.
As far as geometry is concerned, the main ingredient of the G2 space is the associative 3-form $\Phi$, which contains the moduli that control the sizes of the 3-cycles. 
With these ingredients, one can readily write down the real superpotential of the 3D $\mathcal{N}=1$ effective supergravity 
\be \label{Tree level superpotential}
P = \frac{1}{8 {\cal V}^2} \int \left( 
e^{-\phi/2} H_3 \wedge \star \Phi + e^{\phi/4} F_4 \wedge \Phi + e^{5\phi/4} \star F_0 
\right) \,, 
\ee 
where $\cal V$ is the volume of the G2 space and $\phi$ is the dilaton. 
We will analyze the ingredients that enter the superpotential in more detail once we introduce the specific space that we will work with. 
For the moment, we note that the RR fluxes satisfy appropriate Bianchi identities 
\be
dF_p = H_3 \wedge F_{p-2} + J_{p+1}  \,, 
\ee
where $J_{p+1}$ is the current of the D(8-p)-branes/O(8-p)-planes, which we will specify later on. 
As far as the vacuum structure of the effective 3D supergravity is concerned, for real scalar moduli $\varphi^i$, with $i$ running over all moduli, the potential is
\be
V = g^{ij} \partial_i P \partial_j P - 4 P^2 \,.  
\ee
The moduli space metric $g_{ij}(\varphi^i)$ is Riemannian, since the moduli space in 3D $\mathcal{N}=1$ supergravity is real. 
This means that the condition for supersymmetric AdS$_3$ is simply $\partial_i P=0$ with $P\ne 0$.

One way to build G2 spaces is by resolving the singularities of  appropriate orbifolds \cite{Joyce:2002eb}. 
There are various ways to build such orbifolds, and our focus here is on setups with known and tractable desingularizations. 
In addition, we want to have full moduli stabilization in the untwisted sector. 
To this end, we look at a specific orbifold from \cite{Joyce1,Joyce2} which we have found by inspection to work, 
and which is described by the quotient 
\be
\label{ORB}
T^7 / (Z_2)^3 \,. 
\ee
Here the $(Z_2)^3$ group is given by the following three involutions 
\be
\begin{aligned}
\label{Z2s}
\Theta_\alpha : y^i & \to (-y^1, -y^2, -y^3, -y^4, y^5, y^6, y^7) \,, 
\\[0.5mm]
\Theta_\beta : y^i & \to (-y^1, 1/2-y^2, y^3, y^4, -y^5, -y^6, y^7) \,,
\\[0.5mm]
\Theta_\gamma : y^i & \to (1/2-y^1, y^2, -y^3, y^4, -y^5, y^6, -y^7) \,, 
\end{aligned}
\ee
where the $\Theta_i$ acts as generators of the involutions and toroidal coordinates already satisfy the periodic identification $y^i \sim y^i + 1$. 
The desingularization is presented in \cite{Joyce1,Joyce2} and makes use of the Eguchi--Hanson space; we will treat this in the next section. 
By inspection, we have found a specific flux setup that allows full moduli stabilization even in the parametrically large flux limit. 

In \cite{Farakos:2025bwf}, a similar orbifold (differing only in $\Theta_\gamma$) was instead found to lead to integer dual conformal dimensions in the untwisted sector, at the price of a persistent flat direction in the parametric scale separation limit. 
We do not know whether some other choice of fluxes could achieve full moduli stabilization in the setup of \cite{Farakos:2025bwf} at parametrically large volume and weak coupling, but we have not found one. 
We therefore revisit G2 flux compactifications using the orbifold generated by the involutions in \eqref{Z2s}.

The orbifold \eqref{ORB} allows for a specific basis of harmonic forms which are 
\be
\label{3basis}
\phi_i = \left(dy^{136}, dy^{246}, - dy^{235}, dy^{127}, - dy^{567},  - dy^{347}, dy^{145} \right) \ , \quad i = 1 , \dots, 7 \, , 
\ee
and also have a dual basis for harmonic 4-forms given by 
\be
\label{4basis}
\psi_i = \left(dy^{2457}, dy^{1357}, - dy^{1467}, dy^{3456}, - dy^{1234},  -dy^{1256}, dy^{2367} \right) \ , \quad i = 1 , \dots, 7 \, . 
\ee
These forms satisfy the orthogonality condition $\int \phi_i \wedge \psi_j = \delta_{ij}$ and serve as a basis for the fluxes and currents of the sources that enter the construction, which we describe next.

Due to the orbifold, the orientifolds that we can work with are: 
\begin{align}\label{orientifold table}
\begin{pmatrix} 
&{\rm O}2:\quad & \sim & - & - & - & - & - & -  \\
&{\rm O}6_1:\quad & - & \times & - & \times & \times & - & \times  \\ 
&{\rm O}6_2:\quad &\times & - & \times & - & \times & - & \times  \\ 
&{\rm O}6_3:\quad &\times & \sim & - & \times & - & \times & \times \\ 
&{\rm O}6_4:\quad & \sim & \sim & \times & \times & \times & \times & -  
\end{pmatrix} \, . 
\end{align}
Here, we use $\times$ to indicate the internal directions that are parallel to the orientifold planes, 
i.e. directions belonging to their world-volume. 
With the symbol $-$ we indicate directions where the orientifold is localized at the positions $0,1/2$, 
while with $\sim$ we refer to directions where the orientifold is localized at the positions $1/4,3/4$. 
These positions are found by combining the O2 involution with the G2 involutions and identifying the loci of fixed points as the orientifold surfaces (notice that the O2 starts with $\sim$). 
Indeed, the O2 involution is 
\be
\sigma: y^i \to (1/2-y^1, -y^2, -y^3, -y^4, -y^5, -y^6, -y^7) \, . 
\ee
One can check that combining this with the orbifold involutions gives the O6-planes in Table \ref{orientifold table}. 

The D-branes are chosen for tadpole cancellation purposes, which will become clear once we discuss our flux choices. 
The D-branes are of course free to move, and one should determine their positions by extremizing the appropriate superpotential (or some suitable potential) that also includes the D-brane positions. 
However, since their moduli space is compact, they are not expected to cause serious inconsistencies in our construction, 
and we leave the analysis of their full stabilization for future work. 
We therefore indicate their positions with the $=$ symbol.
The D-branes are: 
\begin{align}
\begin{pmatrix} 
&{\rm D}2:\quad & = & = & = & = & = & = & =  \\ 
&{\rm D}6_{5}:\quad & \times & \times & \times & \times & = & = & =  \\ 
&{\rm D}6_{6} :\quad &\times & \times & = & = & \times & \times & =  \\
&{\rm D}6_{7} :\quad & = & \times & \times & = & = & \times & \times  
\end{pmatrix} \, . 
\end{align}
One can see that the D-branes are mutually supersymmetric with the orientifolds. 
The index on the D6 (likewise for the O6) indicates which 4-cycle is wrapped by the object's world-volume, or equivalently which harmonic 3-form describes its current. 
The D2-branes can be chosen to completely cancel the net O2 contribution to the tadpole, 
or to dominate it, leading to a scaling with large D2 number; 
we will analyze both behaviors.

One can also have D6-branes at the O6 positions, wrapping the same 4-cycles as the O6-planes, as long as they do not exceed the latter's contribution to tadpole cancellation. 
Exceeding it would not be a problem per se, 
but scale separation with moduli stabilization may be ruined, since it is typically supported by the negative tension of the orientifolds (see e.g.\ \cite{Gautason:2015tig}). 
The D6-branes can be present both in models where the D2 number scales and in models without D2 scaling.

In the main part of the article we cover the case where the D2 number scales and D6-branes are also present in the flux background. 
Other distinct cases, with different D2 and D6 behavior, are analyzed in the appendix.

\subsection{Tadpole cancellations}
In this section, we show how the tadpoles on the internal manifold cancel for the setup described above. We do this at the level of the integrated Bianchi identities, demanding no net internal charge.

The key ingredients are O6- and O2-planes, D2- and D6-branes, 
the RR and NS-NS fluxes, and the number of insertions of these sources. 
The numbers of O2- and O6-planes on the covering torus are $2^7$ and $2^3$, respectively. 
We recall that the charge of an O-plane in units of the D-brane charge in flat space is 
\begin{equation}
    Q_{\rm Op} = -2^{p-5} Q_{\rm Dp} \,. 
\end{equation}
Looking at the O2 tadpole on the covering torus, we have 
\begin{equation}\label{O2 tadpole}
\int F_4 \wedge H_3 = (2 \pi)^5 \left(128 \times \left(\frac{1}{8 \times 8}\right)- N_{\rm D2}\right) \quad \to \quad \int F_4 \wedge H_3 = (2 \pi)^5 (2- N_{\rm D2}) \,. 
\end{equation}
Notice that tadpole cancellation on the quotient space results in the same effective O2 charge. On the quotient space, the number of O2-planes is reduced to 32 by the identifications under the orbifold involutions.

One would realize that the O2-planes are all localized at points that are identified with each other under the $\mathbb{Z}_2$ involution $\Theta_{\gamma}$. 
This identification reduces the effective charge of these planes by half. 
One can see this by considering $(1,\mathbb{Z}_2)$ as the stabilizer group of the fixed points under the full involution group. Summing over the quotient space then gives an effective charge $Q^{\rm eff}_{\rm O2}= -\frac{1}{16} Q_{\rm D2}$, as opposed to $-\frac{1}{8} Q_{\rm D2}$, which follows from \eqref{O2 tadpole}.

For the O6-planes and the D6-branes, the standard tadpole condition applies. 
We see that net tadpole cancellation is achieved with an appropriate number of D2-branes. 
Next, we present a specific example satisfying the tadpole cancellation described above.

\subsection{Flux example with D6-branes and D2 scaling}\label{Flux example with D6-branes and D2 scali}

Now we can work out the details for the bulk untwisted sector to see how precisely the fluxes are arranged and how moduli stabilization works. 
We have already worked out the tadpole cancellation for a general case and our results also hold here. 
We work with Romans mass given by 
\be
F_0 = m > 0 \ , 
\ee
and the $F_4$ flux is fully isotropic 
\be
\label{F44}
F_4 = - f (\psi_1 + \psi_2 + \psi_3 + \psi_4 + \psi_5 + \psi_6 + \psi_7) \quad , \quad f > 0 \ , 
\ee
while the $H_3$ is essentially non-isotropic because of contributions both from O6-planes and D6-branes 
\be
\label{H33}
H_3 = h (\phi_1 + \phi_2 + \phi_3 + \phi_4) - h (\phi_5 + \phi_6 + \phi_7) \quad , \quad h > 0   \ .  
\ee
This leads to proper D6/O6 tadpole cancellation with an appropriate number of D6-branes taking into account that a D6 charge is a fraction of the O6 charge (with opposite sign). 
For the D2-branes, we have instead 
\be
\label{NND2}
N_{\rm D2} = \frac{hf}{(2 \pi)^5} + 2 \,. 
\ee
Inserting  all the ingredients in the superpotential \eqref{Tree level superpotential}, 
and extremizing with respect to the eight moduli we find that the dilaton/string coupling is 
\be
g_s = e^\phi = 
\frac{14^{3/4} 2 \sqrt{5}}{21}\,
\frac{h}{m^{1/4}f^{3/4}} \,,
\ee
while the physical Einstein-frame periods are
\be
s_{1,2,3,4}
=
\frac{50}{21}\frac{h}{m}e^{-7\phi/4},
\qquad
s_{5,6,7}
=
\frac{50}{3}\frac{h}{m}e^{-7\phi/4} \,.
\ee
One can check that this extremum corresponds to supersymmetric AdS$_3$. 
For the total string-frame volume we find 
\begin{equation}
 {\cal V}_S
 =
 \frac{125\,14^{1/4}\sqrt{5}}{28}
 \left(\frac{f}{m}\right)^{7/4},
\end{equation}
while the Einstein-frame volume is 
\begin{equation}
\label{Ef-volume}
 {\cal V} = \frac{{\cal V}_S}{g_s^{7/4}}
 =
 \frac{3^{7/4}5^{21/8}}{2^{77/16}7^{5/16}}
 \frac{f^{49/16}}{m^{21/16}h^{7/4}}.
\end{equation}
This guarantees parametric large volume (both in Einstein and string frame) and parametrically weak coupling at large $f$. 
The analysis for scale separation carries through exactly as in \cite{Farakos:2020phe} therefore at large $f$ the AdS$_3$ vacuum is scale separated 
\be
\frac{L_{\rm KK}^2}{L_{\rm AdS}^2} \sim f^{-1} \,. 
\ee
Note that the string frame volume scales like $f^{7/4}$ and since the number of D2-branes increases with $f$ we have $\frac{N_{\rm D2}}{{\cal V}_S} \sim f^{-3/4}$, 
therefore they are diluted in the scale separation limit. 
This model exhibits full moduli stabilization, as all masses are nonzero. None of the dual operator dimensions are integers; they are given by (up to higher order decimals for the last three of them) 
\begin{equation}
\Delta
=
\left\{
\frac{34}{25},
\frac{34}{25},
\frac{34}{25},
\frac{112}{25},
\frac{112}{25},
2.89,
4.15,
7.91 
\right\} \,. \end{equation}

\section{Twisted sector moduli stabilization}\label{Section 3: Twisted sector moduli stabilization}

Our aim here is to perform the next step in the development of the general orbifold models presented in \cite{Farakos:2020phe}. 
As shown in subsection \ref{Flux example with D6-branes and D2 scali}, the orbifold construction we are working with allows for full classical moduli stabilization in the untwisted sector. 
In this section, we extend our analysis to the twisted sector of the theory, which arises after a resolution of the singular geometry.\footnote{Typically we think of some of the states of the string twisted sector on singular spaces as being {\it geometrized} into blow-up moduli on a smooth manifold.} 
On top of the fully stabilized background, we treat the twisted sector as a small perturbation. 
A similar analysis has been performed for the DGKT solution \cite{DeWolfe:2005uu}. 
Crucially, it was shown in \cite{Junghans:2023yue} (for DGKT) that upon resolution of the singularities, the O6-planes wrapping the internal cycles no longer intersect. For type IIB solutions, this was recently investigated in the context of non-abelian orbifolds in \cite{Revello:2026eqp}, for a singularity structure similar to ours.
We present a similar analysis in this section and show that twisted-sector moduli stabilization in our models has a similar fate as in DGKT, while in the next section we address the resolution of O6-plane intersections. 
Before turning to the stabilization, let us look at the origin and some features of the twisted-sector moduli.

\subsection{Blow-up moduli}\label{blow-up moduli} 

Here, we will have a closer look at the orbifold and its singularity structure. 
We will first show some key features associated to the resolution of orbifold singularities and familiarize ourselves with the conventions and notations used throughout the rest of the paper.

Let us recall that the orbifold group is generated by $\Theta_\alpha,\Theta_\beta,\Theta_\gamma$, defined in eq.~\eqref{Z2s}. Each involution generates a number of fixed points on the internal space.
Consider $\Theta_\alpha$ to begin with. It generates 16 fixed points on the torus in the $y_1,y_2,y_3,y_4$ directions, leaving the coordinates $y_5,y_6,y_7$ untouched. Hence, the singularity structure close to each fixed point is of the form $T^4/\mathbb{Z}_2 \times T^3$.
Similarly, the other involutions generate 16 fixed points each, giving a total of 48. However, some of these fixed points are mapped onto one another by the shifts in the $\Theta$ generators apparent in eq.~\eqref{Z2s}. These shifts reduce the number of distinct fixed points. One can show that there are in total 8 fixed points of $\alpha ,\beta$ type and 8 fixed points of $\gamma$ type. Each of the former 8 can be locally expressed in complex coordinates as $\mathbb{C}^2/\mathbb{Z}_2 \times T^3$. 
However, the fixed points of $\gamma$ are left invariant under the $\Theta_\alpha \Theta_\beta$ involution 
and hence the singularity structure corresponds to  $\frac{\mathbb{C}^2/\mathbb{Z}_2 \times T^3}{\mathbb{Z}_2}$. 
It is well known that singularities of this type can be resolved by a standard procedure in which a ball of fixed radius around each fixed point is replaced by a smooth Euclidean manifold \cite{Gibbons1979,Joyce:2002eb,Joyce1,Joyce2}. 
In particular, one has two possible resolutions for the $\gamma$ fixed points differing by the action of residual $\mathbb{Z}_2$ on the resolved space. 
Following standard nomenclature we call these topologically different resolution choices to be of the first and second type.

The underlying complex manifold of the Eguchi--Hanson space is the minimal
crepant resolution of $\mathbb{C}^2/\mathbb{Z}_2$. It can be described
equivalently as the cotangent bundle of the two-sphere
\begin{equation}
    X_{\mathrm{EH}} \cong T^* S^2 \cong T^*\mathbb{CP}^1 \,. 
\end{equation}
The corresponding bundle projection is
\begin{equation}
X_{\mathrm{EH}}
\xrightarrow{\ \pi\ }
\mathbb{CP}^1,
\end{equation}
where the fiber over every point of $\mathbb{CP}^1$ is one copy of
$\mathbb{C}$. Locally, a point of $X_{\mathrm{EH}}$ may therefore be
described by coordinates $(w,t)$, where $w$ parametrizes the base
$\mathbb{CP}^1$ and $t$ parametrizes the cotangent fiber. The zero section is the embedding
\begin{equation}
i:
E\simeq\mathbb{CP}^1
\hookrightarrow
X_{\mathrm{EH}},
\qquad
w \longmapsto (w,0).
\end{equation}
The sphere $E$ is called the exceptional curve, exceptional divisor, or bolt. For a more elaborate discussion and metric structure of the Eguchi--Hanson space, see Appendix~\ref{Appendix Eguchi Hanson}.
One essentially replaces the inside of the ball ($B^4$) around the fixed point by a truncated Eguchi--Hanson space, and the boundary structure of both being $S^3/\mathbb{Z}_2$ facilitates a proper gluing. One then forms a product of this truncated space with a fixed $T^3$. Applying this surgical procedure to all the fixed points of this type will resolve the singular G2 into a smooth G2. 
Symbolically,
\begin{equation}
    B^4/\mathbb{Z}_2 \times T^3 \rightarrow X_{EH} \times T^3.
\end{equation}
This resolution inevitably changes the topology of the manifold. The exceptional sphere ($\mathbb{CP}^1$) supplies one new two-cycle 
\begin{equation}
    E \in H_2(X_7).
\end{equation}
Taking its product with each tangent one-cycle of the fixed torus gives three new three-cycles 
\begin{equation}\label{New periods}
    \chi_a=E \times S_{y^a}^1.
\end{equation}

As mentioned earlier, for the Joyce orbifold under discussion there are four inequivalent singular tori of each type $\alpha, \beta$ and eight of $\gamma$ type, making it 16 in total. Resolving all of them would make the total \cite{Joyce2} 
\begin{equation}\label{New betti numbers}
 b_2= 8+l , \quad b_3= 47-l , \qquad l=0,1...8 \ , 
\end{equation} 
where $l$ is the number of involutions of the first type for the $\gamma$ fixed points. 
We will see in Sec.~\ref{Counting the active ingredients} how the orientifold action naturally forces upon us the second kind of involution for these points.
It is clear that consequently there is also an increase in the number of $b_4$ and $b_5$ Betti numbers. 
For the moment we will perform a general analysis keeping the naive counting of the Betti numbers and postpone including the orientifold projections until Sec.~\ref{Counting the active ingredients}.

The main takeaway from this section is that the resolution of the fixed points can be carried out locally for each fixed point and consequently one obtains a smooth G2 manifold. Furthermore, the exceptional sphere along with the transverse 3-tori supplies the smooth manifold with extra two- and three-cycles. 
The presence of these new blow-up cycles will potentially lead to the presence of new axion fields (from the $B_2$ and the RR fields), 
which are in principle fields that do not control the radii (therefore neither the scale separation). 
Since these fields have a compact moduli space, one expects them to be stabilized at some value compatible with supersymmetry. 
This is an interesting check that can be performed once we have the full superpotential for the full blow-up moduli sector, which we leave for future work. 
Our main interest here is instead to verify that the actual blow-up size moduli can be maintained at finite size.

\subsection{Stabilization of blow-up modes}\label{Stabilization of blow-up modes}

Now we analyze the metric moduli introduced by resolving the orbifold singularities. 
The singular orbifold contains seven untwisted bulk moduli, 
associated with deformations of the invariant closed three-forms defining the G2 structure. 
After resolving the singular components, 
the smooth manifold additionally contains $32$ blow-up moduli, while the orientifold implications for the counting are considered only in Sec~\ref{Counting the active ingredients}.
For this section and Sec.~\ref{Bosonic reduction cross-check}, we will assume that \(a=1,\ldots,N\) labels the blow-up moduli corresponding to each fixed point. We know that $\mathcal{N}$ is at most three but could differ depending on details, 
hence for our purpose we take $a=1,2,3$ as we only intend to provide the general idea for their stabilization. 

It is important to note that these three moduli should not be interpreted as three independent sizes of the exceptional sphere. 
For each resolved component, denoted by $(\tau,n)$, they form a three-component resolution vector $u^{(\tau,n,a)}$, with $a=1,2,3$. 
Away from the orbifold point, this vector can equivalently be parametrized by one magnitude, controlling the size $\rho_{(\tau,n)}$ of the Eguchi--Hanson bolt, and a unit vector $n^a_{(\tau,n)}\in S^2$, whose two independent angles describe the orientation of the Eguchi--Hanson hyperk\"ahler structure relative to the bulk G2 structure. Thus, the three blow-up fields are naturally viewed as one radial and two angular deformation modes. 
We elaborate on this geometrical interpretation in Appendix~\ref{Appendix Eguchi Hanson}, 
while for the purposes of this section we will only be concerned with stabilizing them. 
These bulk and blow-up moduli constitute our metric moduli, with $s_i$ denoting the former and $u_a$ the latter. 
In general, metric moduli correspond to fluctuations of the metric, which can be written in terms of the periods of the three-form defining the G2 structure. 
The Eguchi--Hanson space contains one exceptional two-cycle $E_{\tau,n}\simeq S^2$ and one associated normalizable harmonic two form $\nu_{\tau,n}$. 
Taking the product of $\nu_{\tau,n}$ with each of the three one-cycles of the fixed three-torus gives localized three-forms, 
schematically of the form 
\begin{equation}\label{exceptional forms}
    \zeta_{(\tau,n,a)}
    \simeq
    \nu_{\tau,n}\wedge d\xi^a \, .
\end{equation}
The form of $\nu_{\tau,n}$ is given in \eqref{2 form} and $\xi^a$ denotes the one-form on the tangent torii whose sizes are denoted by $R_{\tau}^a$. 
The blow-up moduli are defined as the periods of the full associative three-form over the exceptional three-cycles with appropriate normalization 
\begin{equation}\label{Periods relation}
    u^{(\tau,n,a)} =
    \frac{1}{8}
    \int_{\chi^{(\tau,n,a)}}\Phi_{EH} =
    -\frac{\pi}{2}
    R^a_{\tau}\rho^2_{(\tau,n)}
    n^a_{(\tau,n)} \quad ,
    \quad
    \sum_{a=1}^{{3}}
    \left(n^a_{(\tau,n)}\right)^2=1 \ .  
\end{equation}
Here $\rho_{(\tau,n)}$ denotes the size of the exceptional sphere for a fixed point $n$ generated by the $\Theta_\tau$ involution. $\Phi_{EH}$ here is taken as the relevant part of $\Phi_{\mathrm{loc}}$ defined in eq.~\ref{LocalPhi} which detects a given exceptional cycle $\chi^{(\tau,n,a)}$.
Equation~\eqref{Periods relation} makes explicit that the three periods 
$u^{(\tau,n,a)}$ are Cartesian coordinates on the local blow-up deformation space. 
Indeed, they determine the physical size through 
\begin{equation}
    \rho_{(\tau,n)}^4
    =
    \frac{4}{\pi^2}
    \sum_{a=1}^{3}
    \frac{\left(u^{(\tau,n,a)}\right)^2}
         {\left(R^a_{\tau}\right)^2} \, , 
\end{equation}
while their direction determines the two orientation angles given by 
\begin{equation}
    n^a_{(\tau,n)}
    =
    -\frac{2u^{(\tau,n,a)}}
    {\pi R^a_{\tau}\rho_{(\tau,n)}^2}.
\end{equation}
At the orbifold point, where all three periods vanish, the exceptional
sphere collapses and its orientation is no longer defined.
One can calculate the volume of the smooth manifold in Einstein frame with respect to the singular manifold to leading order in the blow-up radius as was done in \cite{Lukas:2003dn}. 
To this end we have 
\be 
    \mathcal V = V_0\,\Omega(u),
    \quad 
    V_0 = \left(\prod_i s^i\right)^{1/3},
\ee 
where 
\be    
    \Omega(u) =
    1-\frac{8}{3}
    \sum_{\tau,n}
    \sum_{a=1}^{3}
    \frac{\left(u^{(\tau,n,a)}\right)^2}
         {s^{A(\tau,a)}s^{B(\tau,a)}}
    +{\cal O}(u^4).
\ee 
For a fixed component of type $\alpha$, the relevant pairs in our
conventions are\footnote{Notice the difference in convention from \cite{Lukas:2003dn}, mainly due to the labeling of G2 forms.}
\begin{equation}
    \bigl(A(\alpha,a),B(\alpha,a)\bigr)
    =
    (5,7),\ (3,4),\ (1,6),
    \qquad a=1,2,3 .
\end{equation}
Notice that blowing up the singularity unavoidably also affects the size of the associated bulk cycles, 
which corresponds to (small) shifts in the value of the bulk moduli. 
We will revisit this point in different contexts throughout. It is also easy to observe that the limit $u_a \to 0$ takes us back to the singular orbifold case. For the bulk moduli associated with the blow-up moduli of the
$\alpha$-, $\beta$-, and $\gamma$-type fixed points, we refer the reader
to table~\ref{tab:bulk-moduli-associations}.

\begin{table}[t!]
\centering
\begin{tabular}{@{}llll@{}}
\toprule
Singular locus $\tau$
& $a=1$
& $a=2$
& $a=3$ \\
\midrule
$F_\alpha$
& $(5,7)$
& $(3,4)$
& $(1,6)$ \\
$F_\beta$
& $(1,2)$
& $(1,4)$
& $(3,6)$ \\
$F_\gamma$
& $(5,6)$
& $(1,7)$
& $(3,2)$ \\
\bottomrule
\end{tabular}
\caption{Pairs of bulk-modulus indices associated with the three
blow-up moduli $u_a$ at each type of singular locus.}
\label{tab:bulk-moduli-associations}
\end{table}
To stabilize the untwisted and exceptional moduli, we introduce mutually
dual bases of the third and fourth cohomology class. We write
\begin{equation}
\label{eta-lambda}
    \eta_I=(\phi_i,\zeta_a)\in H^3(X_7),
    \qquad
    \lambda^I=(\psi^i,\tilde\zeta^{a})\in H^4(X_7),
\end{equation}
as the bases that are normalized by the topological
Poincar\'e pairing
\begin{equation}
\label{Topological linking}
    \int_{X_7}\eta_I\wedge\lambda^J
    =
    \delta_I{}^J.
\end{equation}
In particular,
\begin{equation}
    \int_{X_7}\phi_i\wedge\psi^j=\delta_i{}^j,
    \qquad
    \int_{X_7}\zeta_a\wedge\tilde\zeta^b
    =\delta_a{}^b,
\end{equation}
while the bulk-exceptional cross-pairings vanish in this dual basis. 
The cohomology class of the associative three-form can then be expanded as
\begin{equation}
\label{ExpansionIII}
    [\Phi]
    =
    s^i[\phi_i]
    +
    u^A[\zeta_A].
\end{equation}
Here $s^i$ are the untwisted metric moduli, while $u^A$ are the exceptional moduli.
Equation~\eqref{ExpansionIII} is to be thought of as global cohomological expansion not a pointwise description of $\Phi$ defining the full G2. It only states that the single global cohomology class \([\Phi]\)  inherits bulk and exceptional period directions. In the bulk region, the associative form is approximated by the
orbifold expression $\Phi_0(s)$, whereas in a resolved neighborhood $U$
it is represented by the smooth local form
$\Phi_{\mathrm{loc}}^{(A)}(s,u)$. 
Therefore we have 
\begin{equation}
    \Phi\big|_{U_{\mathrm{bulk}}}
    \simeq
    \Phi_0(s) \ , 
    \quad
    \Phi\big|_{U_{\mathrm{loc}}}
    \simeq
    \Phi_{\mathrm{loc}}(s,u) \, .
\end{equation}
These are not independent contributions to $\Phi$, 
but local representations of the same globally defined G2 structure, 
matched in their overlap regions. 
We retain $H_3$ flux only in the untwisted sector, while allowing both bulk
and exceptional components of $F_4$ similarly representing them in terms of expansion over elements of cohomology class as follows 
\begin{equation}
\label{ExpansionII}
    H_3=h_i[\phi_i],
    \qquad
    F_4=e_i[\psi^i]+q_A[\tilde\zeta^A].
\end{equation}
The coefficients $h_i$ and $e_i$ are bulk flux quanta, while $q_A$ denote
the exceptional $F_4$ flux quanta which we denote as $F_4^{ex}$.
The moduli for the blow-ups still remain real consistent with the real formulation of the 3D $\mathcal{N}=1$  supergravity. 
We therefore expand the tree level superpotential in \eqref{Tree level superpotential} up to quadratic order in the blow-up moduli. 
The substitution of \eqref{ExpansionIII}, \eqref{ExpansionII} in \eqref{Tree level superpotential} has to be also followed with proper treatment of the terms containing the NS-NS fluxes, which contains the Hodge star with respect to the smooth G2 metric as the Hodge star depends on the full metric after the blow-up. 
After the expansion of the superpotential to quadratic order in the $u_a$ blow-up moduli we get 
\begin{equation}
\label{Superpotential correction}
\begin{aligned}
P \ = &\ \frac{F_0}{8} {V_0}^{-1} e^{5 \phi / 4}+\frac{1}{8} V_0^{-1} e^{-\phi / 2} \frac{h^i}{s_i}
\\
   & \  +\frac{1}{8} V_0^{-2} e^{\phi / 4}\left(s^i e_i+\sum_{\tau ,n,a}u_a q_a\right)+\frac{1}{2}\sum_{\tau ,n,a} X_a u_a^2+{\cal O}\left(u^4\right) \,, 
\end{aligned}
\end{equation}
where 
\be 
X_a=d_a\left\{\frac{2}{3} F_0 e^{5 \phi / 4} V_0^{-1}+e^{-\phi / 2} V_0^{-1}\left[\frac{2}{3} \mathcal{H}+2\left(H_{A_a}+H_{B_a}\right)\right]+\frac{4}{3} e^{\phi / 4} V_0^{-2} f\right\} \,, 
\ee 
with $H_i \equiv h^i / s^i$, $\mathcal{H} \equiv \sum_i H_i$ and $f \equiv \sum_i s^i e_i$. 
Notice the expression contains terms linear in $u_a$, and $X_a$ denotes the coefficient of the quadratic order contribution. 

The three supersymmetry conditions are independent equations given by 
\begin{equation}
    0
    =
    \frac{\partial P}
         {\partial u^{(\tau,n,a)}}
    =
    L_{(\tau,n,a)}
    +
    X_{(\tau,n,a)}u^{(\tau,n,a)}
    +{\cal O}(u^3),
    \qquad a=1,2,3 ,
\end{equation}
and, provided the quadratic form is non-degenerate, they fix all three components 
\begin{equation}
    u^{(\tau,n,a)}_\star
    =
    -\frac{L_{(\tau,n,a)}}{X_{(\tau,n,a)}}
    + {\cal O} \left(\frac{q^3}{f^3}\right).
\end{equation}
In the polar description of the deformation space, these three equations
fix both the magnitude of the resolution vector, and hence the bolt size
$\rho_{(\tau,n)}$, and its direction $n^a_{(\tau,n)}$, and hence the two
orientation moduli.
With this under control all that remains is plugging the value for the particular choice of fluxes and extremizing the superpotential with respect to all the moduli fields. 
We show general properties of the blow-up moduli stabilization for the specific working model in this part of the article, 
leaving more details for subsection \ref{A specific example} where we also take into account the orientifold truncation on the blow-up modes. 
We leave the discussion of blow-up moduli stabilization of other flux models for the appendix.

For the flux choices we have here (but essentially also generally for G2-orientifold models \cite{Farakos:2020phe}), 
working in Einstein frame, the volume and string coupling in the solution are 
\be 
V_0  \sim\frac{f^{49/16}}{m^{21/16}h^{7/4}} \quad , \quad  g_s \sim \frac{h}{m^{1/4} f^{3/4}} \ . 
\ee 
Substituting the value of the bulk moduli we see that $d_a$ scales like 
\begin{equation}
    d_a \sim \frac{1}{s_as_b} \sim \frac{m^{9/8} h^{3/2}}{f^{21/8}} \,. 
\end{equation}
Correspondingly all the terms inside $X_a$ namely the Romans mass contribution, NS-NS flux and RR flux contribution scale as $h^3 m^2/f^4$. 
Therefore, barring the numerics, each term labeled by $X_a$ goes like 
\begin{equation}
X_a \sim
\frac{h^{9/2}m^{25/8}}{f^{53/8}}.
\end{equation}
The superpotential now looks schematically as 
\begin{equation} 
P=
P_{0}
+
\sum_{a=1}^3L_au_a
+
\frac{1}{2}\sum_{a=1}^3 X_a u_a^2 + \text{higher order terms} \,. 
\end{equation}
The extremization of the superpotential with respect to the blow-up moduli fixes their scalings, 
and with appropriate normalization with respect to bulk moduli we get  
\begin{equation}{\label{superpotential scaling}}
    \frac{u^*_a}{\sqrt{s_as_b}} \sim \frac{q_a}{f} \,. 
\end{equation}
We see that the scale of the blow-up moduli is controlled by the exceptional flux parameter $q_a$ and is smaller than the bulk moduli at large $f$ for appropriate choice of flux satisfying 
\begin{equation}
    q_a \ll f \,. 
\end{equation}
Integrating out the blow-up moduli one gets the corrected superpotential that includes now the blow-up backreaction for the bulk moduli 
\begin{equation} 
    P= P_0 + \Delta P_{*} \quad , \quad \Delta P_{*} \sim {\cal O}(q_a^2/f^2) \,, 
\end{equation}
which essentially yields corrections to the bulk moduli of order 
\begin{equation}
\frac{\delta s^i}{s^i} \sim  \delta\phi
\sim 
{\cal O} \left( q_a^2 / f^2\right) \, . 
\end{equation}
This shows that the correction to the expectation values of bulk moduli is small and the backreaction of the blow-up moduli can be neglected at leading order in perturbation at large $f$ as long as $q_a^2 \ll f^2$. 
We also emphasize that higher order terms $O(u^4)$ get even smaller parametrically with large $f$ and can thus further be neglected safely.

Another important condition that this stabilization achieves is, at each order in perturbation we remain in the large volume regime, where the blow-up and bulk scales are larger than the string scale, keeping us in the desired regime of string theory where supergravity is a good approximation. 
We show this by observing that the characteristic bulk radius scales at leading order as (restoring momentarily the string length) 
\begin{equation}
\frac{R_I}{\ell_s}
\sim
\left(\frac{f}{m}\right)^{1/4}.
\end{equation}
Using the standard Eguchi--Hanson period relation in eq.\eqref{Periods relation}, the exceptional cycle scale behaves as 
\begin{equation}
\frac{\rho^2}{\alpha'}
\sim
\frac{Q_{\mathrm{eff}}}{\sqrt{fm}},
\end{equation}
where $Q_{\mathrm{eff}}$ is the appropriate shape-weighted norm of the
exceptional fluxes. We will define it more carefully in the next subsection with a cleaner example. 
For a fixed $m$, a parametrically controlled resolved regime therefore requires
\begin{equation}
\sqrt{f}\ll Q_{\mathrm{eff}}\ll f.
\end{equation}
This can be achieved by fine tuning the dependence of exceptional flux $q_a$ with respect to the bulk $F_4$ scalings.

In subsection \ref{A specific example}, once we have determined the precise orientifold positions after the resolution, 
we will further scrutinize our working example. 
We will see that the heuristic argument works the way we presented it above with no problematic cancellations in the numerator or the denominator.

\subsection{Bosonic reduction analysis} \label{Bosonic reduction cross-check}

In this section, we illustrate the stabilization mechanism for the blow-up moduli working directly with the bosonic action, 
focusing only on the $F_4$ part. 
This provides a useful cross-check that the dependence obtained from the superpotential is already present in the dimensionally reduced ten-dimensional action. 
As a byproduct, 
we will see that the actual bosonic-sector mechanism behind the blow-up moduli stabilization here is different compared to DGKT \cite{DeWolfe:2005uu}.

Let us again focus only at one particular blow-up fixed by the generator of involution $\Theta_{\alpha}$, we basically drop the indices $\tau,n$ from the blow-up moduli $u_a$.This component contributes three blow-up moduli $u_a$, with $a=1,2,3$. Each $u_a$ couples through the corrected volume to a definite pair of bulk moduli,
\begin{equation} 
\left(A_a,B_a\right) = (5,7),\,(3,4),\,(1,6), \qquad a=1,2,3 .
\end{equation}
Thus, $s^{A_a}$ and $s^{B_a}$ denote the two bulk moduli associated with the component $u_a$.\footnote{The extension to include all fix points has no obstruction in principle but is considerably more tedious and messy.}

We emphasize that restricting the calculation to the $F_4$ sector does not by itself constitute a complete proof of perturbative stability. However, our purpose here is instead to illustrate how the blow-up moduli stabilization works in the bosonic sector and verify their parametric scaling. 
The rest of the bosonic sector exhibits the same scalings, which is guaranteed by the supersymmetry, 
and are hence neglected as we are not interested in the numerics here. 
If our full stabilization argument was based solely on the bosonic sector we would need to go through all the cumbersome details, 
but here we are merely interested in extracting the behavior of the bosonic sector, 
while we know that stabilization does take place due to the superpotential analysis. 
The relevant term in the ten-dimensional Einstein-frame action therefore is 
\begin{equation} 
S_{F_4} = -\frac{1}{4\kappa_{10}^2} \int e^{\phi/2}F_4\wedge\star_{10}F_4 \, . 
\end{equation}
Working in the same setup as we did for the superpotential analysis we introduce mutually dual bases of the third and fourth cohomology \eqref{eta-lambda}, 
which are normalized according to \eqref{Topological linking}. 
This is the topological Poincar\'e-duality pairing which should not be confused with the metric-dependent Hodge inner product introduced below.

The $F_4$ is expanded in the basis of the harmonic 4 forms as \eqref{ExpansionII}, 
therefore in the total four-form basis $(\psi^i,\tilde{\zeta}^a)$ the flux vector is 
\begin{equation} 
\label{NI}
N_I=(e_i,q_a) \, . 
\end{equation} 
For this particular blow-up the associated bulk moduli are cycles $(s^a,s^b)$. 
We define the Hodge matrix
\begin{equation}  
K_{IJ} = \frac{1}{4\mathcal V} \int_{X_7}\eta_I\wedge\star_7\eta_J . \end{equation}
Unlike the Poincar\'e-duality pairing, $K_{IJ}$ depends on the internal metric. 
In particular, after resolution, the bulk and exceptional representatives need not be Hodge-orthogonal, that is 
\begin{equation} \int_{X_7}\phi_i\wedge\star_7\zeta_a \neq 0 . \end{equation} 
This bulk-exceptional Hodge mixing will be responsible for the terms linear in the blow-up moduli. 
Using the duality between the bases, one finds \begin{equation} \int_{X_7}\lambda^I\wedge\star_7\lambda^J = \frac{1}{4\mathcal V} \left( K^{-1}\right)^{IJ}.
\end{equation}
Reduction to three dimensions, followed by the Weyl rescaling to the three-dimensional Einstein frame, therefore gives 
\begin{equation}\label{3d bosonic potential}
    V^{(3d)}_{F_4}= \frac{1}{2}e^{\phi/2}\mathcal{V}^{-3} \int F_4 \wedge \star F_4 = \frac{1}{8} e^{\phi/2}\mathcal{V}^{-4} N_I K^{IJ}N_J \,. 
\end{equation}
For one component $u_a$, the relevant fields are ordered as 
\begin{equation} 
\left(u_a,s^{A_a},s^{B_a}\right) \,. 
\end{equation}
Suppressing now the label $a$ temporarily, the corresponding sub-block of the Hodge matrix is deduced from the standard K\"ahler metric relation 
\begin{equation}
    K_{I  J}= \frac{-3}{4} \partial_{I} \partial_{J} \ln{\mathcal{V}} \,, 
\end{equation}
and gives 
the corresponding sub-block of the Hodge matrix 
\begin{equation} \label{Hodge subblock} K_{\left(u,A,B\right)} = \begin{pmatrix} \dfrac{4}{s^As^B} & -\dfrac{4u}{(s^A)^2s^B} & -\dfrac{4u}{s^A(s^B)^2} \\[2mm] -\dfrac{4u}{(s^A)^2s^B} & \dfrac{1}{4(s^A)^2} +\dfrac{4u^2}{(s^A)^3s^B} & \dfrac{2u^2}{(s^A)^2(s^B)^2} \\[2mm] -\dfrac{4u}{s^A(s^B)^2} & \dfrac{2u^2}{(s^A)^2(s^B)^2} & \dfrac{1}{4(s^B)^2} +\dfrac{4u^2}{s^A(s^B)^3} \end{pmatrix} +\mathcal O(u^3). 
\end{equation}
Its inverse, to the order required for our calculation, is
\begin{equation} \label{inverse Hodge subblock} K_{\left(u,A,B\right)}^{-1} = \begin{pmatrix} \dfrac{s^As^B}{4}+\mathcal O(u^2) & 4s^Au & 4s^Bu \\[2mm] 4s^Au & 4(s^A)^2 & 32u^2 \\[2mm] 4s^Bu & 32u^2 & 4(s^B)^2 \end{pmatrix} +\mathcal O(u^3).
\end{equation} 
The hidden correction of order $u^2$ in the first entry contributes only at order $q^2u^2$ and will not be needed for the leading small-$q$ stabilization mechanism.

For the three blow-up components associated with $\Theta_\alpha$, the relevant part of the Hodge matrix can be displayed more transparently as \begin{equation} \begin{aligned} K_{\text{rel}} &= K_{\left(u_1,s^5,s^7\right)} \oplus K_{\left(u_2,s^3,s^4\right)} \oplus K_{\left(u_3,s^1,s^6\right)} \oplus K_{\text{spectator}} +\mathcal O(u_au_b). \end{aligned} \end{equation}
At the orbifold point, $u_a=0$, the off-diagonal entries vanish and one recovers the original diagonal bulk block together with a decoupled exceptional norm. The important new feature away from the orbifold point is the appearance of the off-diagonal entries proportional to $u_a$.\footnote{ We note that blow-ups at fixed points associated with different generators do not affect the entries corresponding to this particular fixed point, justifying our treatment of the fixed points independently.} 
In contrast, 
one can easily check that the diagonal term of the blow-up cycle is independent of its scaling, 
therefore it can not oppose a shrinking blow-up cycle and help stabilize it. 
The main difference therefore as opposed to the procedure known in the literature for DGKT vacua (Section 3.5 in \cite{DeWolfe:2005uu}) is the presence of off diagonal terms in the above expression \eqref{inverse Hodge subblock}. 

Substituting the expressions \eqref{inverse Hodge subblock} and \eqref{NI} into \eqref{3d bosonic potential} gives 
\begin{equation} 
\label{eq:VF4expanded} 
V_{F_4}  
= V_{F_4}^{(0)} + e^{\phi/2}V_0^{-4} 
\left[ 
\frac{s^{A_a}s^{B_a}}{32}q_a^2 + q_aC_au_a 
+ \left( 8D_a + \frac{16\Sigma_e} {3s^{A_a}s^{B_a}} \right)u_a^2 
\right] 
+ {\cal O} \left(q_a^2u_a^2,u_a^3\right), 
\end{equation} 
where 
\begin{equation} 
V_{F_4}^{(0)} = \frac{1}{2}e^{\phi/2}V_0^{-4}\Sigma_e \,, \qquad \text{with}\quad  \Sigma_e= \sum_{i} e_i^2 s_i^2 \,, 
\end{equation} 
is the potential evaluated on the unresolved bulk background. 
Here, $C_a$ and $D_a$ denote the coefficients of the linear and quadratic contributions in $u_a$ to the potential, respectively, with
\begin{equation}
 \qquad C_a=e_{A_a}s^{A_a}+e_{B_a}s^{B_a},\qquad D_a=e_{A_a} e_{B_a}.
\end{equation}

Equation~\eqref{eq:VF4expanded} makes the stabilization mechanism transparent. The diagonal exceptional flux contribution proportional to $q_a^2$ is independent of $u_a$ at the displayed order and therefore does not by itself displace the field from the orbifold point - it is the off-diagonal terms that do the job which are proportional to $q_a$. 
To illustrate what happens if we include the remaining bosonic sector contributions, 
let the potential near the $q_a=0$ bulk solution take the generic form 
\begin{equation} \label{local blowup potential} V = V^{(0)} + J_au_a + \Lambda_au_a^2 + \mathcal O\left(q_a^2u_a^2,u_a^3\right), \end{equation} 
where the $F_4$ contribution to the linear coefficient is \begin{equation} J_a = e^{\phi_0/2}V_0^{-4}q_aC_a . \end{equation} 
Then the coefficient $\Lambda_a$ receives contributions from all sectors of the bosonic action. 
The remaining terms can modify its numerical value, but on the bulk solutions considered above they occur at the same parametric order as the $F_4$ contribution. Provided that $\Lambda_a$ is nonzero, extremization gives 
\begin{equation} 
\label{eq:ustar} u_{a\star} = -\frac{J_a}{2\Lambda_a} + \mathcal O(q_a^2). 
\end{equation} 
Of course  $\Lambda_a$ being nonzero is granted assuming one gets an extremization from the superpotential analysis - we will show this to hold for our working example explicitly later on. 
If one retains only the contribution displayed in \eqref{eq:VF4expanded}, 
this becomes 
\begin{equation} 
u_{a\star} = -\frac{q_aC_a} {2\left( 8D_a+ \dfrac{16\Sigma_e} {3s^{A_a}s^{B_a}} \right)} + \mathcal O(q_a^2). 
\end{equation} 
For the flux choice in \ref{Flux example with D6-branes and D2 scali}, $e_i=-f$, and using the background scalings obtained above, one has parametrically \begin{equation} 
J_a \sim e^{\phi_0/2}V_0^{-4} q_af\sqrt{s^{A_a}s^{B_a}}, \qquad \Lambda_a \sim e^{\phi_0/2}V_0^{-4}f^2 . 
\end{equation} 
Consequently we have 
\begin{equation} 
\label{eq:bosonic scaling} 
\frac{u_{a\star}} {\sqrt{s^{A_a}s^{B_a}}} \sim \frac{q_a}{f}.
\end{equation}
This agrees with the scaling derived from the superpotential in \eqref{superpotential scaling}. 
By choosing the exceptional fluxes parametrically smaller than the bulk flux, 
the resolution can therefore be kept small relative to the bulk scales while remaining away from the orbifold point.

The distinction from the familiar DGKT mechanism can now be stated more precisely. The localized harmonic two-form on Eguchi--Hanson is middle-dimensional in the four-dimensional Eguchi--Hanson space. Its norm is therefore invariant under an overall rescaling of the local metric. Consequently, the leading diagonal exceptional-flux term is independent of the blow-up size and cannot by itself produce a parametrically small nonzero resolution. In the present construction, it is the bulk--exceptional Hodge mixing, together with the quadratic terms inherited from the bulk flux energy and the corrected volume, that displaces and stabilizes the blow-up fields.

\section{Impact of the resolution}\label{Section 4 :Impact of the resolution}

In this section we address the important question of whether orientifold planes intersect in our setups. 
At the singular fixed points of the orbifold the configuration from Table \ref{orientifold table} shows 
that different orientifold planes intersect. 
Intersection of such objects could be regarded as a sign of pathology in the microscopic theory, 
as solutions of intersecting localized sources in supergravity are poorly understood (see for example \cite{Bardzell:2024anh}). 
In this section our main result is that the apparent intersection of these orientifold planes is cured by the resolution procedure described in the previous section. Since the orientifolds intersect on orbifold fixed points it is natural to think that such intersections are eliminated once the fixed points are excised. However, the precise locus of these planes after resolution requires a more involved analysis for a satisfactory explanation of this claim.
We explicitly show that the orientifolds sit on non-self-intersecting surfaces on the blown up exceptional sphere. Our procedure is equivalent to the one for the DGKT vacua done in \cite{Junghans:2023yue}. 
On top of this we will also highlight the non-trivial impact of the resolution on tadpole cancellation.

\subsection{Orientifold intersections}

We recall that the known resolution for a singularity of $\mathbb{C}^2/\mathbb{Z}_2$ is provided by the Eguchi--Hanson space which is locally patched up to the sphere cut out around the singular point. 
To show that the orientifolds do not intersect/self-intersect, 
we zoom into how they each sit on the relevant Eguchi--Hanson space by defining local coordinates on each singular patch and tracking their locus. 
To estimate what type of intersections we have to treat, 
let us briefly mention the orbifold points generated by different involutions and the type of orientifolds that intersect on those points, 
at least in the singular limit. 

Of all the different involutions and their products we see that only $\Theta_\alpha$, $\Theta_\beta$, and $\Theta_\gamma$ have fixed sets.
Their fixed loci on the covering torus are
\be 
F_\alpha: 
y^1,y^2,y^3,y^4\in\left\{0,\frac12\right\},
\qquad
(y^5,y^6,y^7)\ \text{free},
\label{eq:F-alpha}
\ee
for the $\Theta_\alpha$, 
then we have 
\be
F_\beta:
y^1,y^5,y^6\in\left\{0,\frac12\right\},
\qquad
y^2\in\left\{\frac14,\frac34\right\},
\qquad
(y^3,y^4,y^7)\ \text{free},
\label{eq:F-beta}
\ee
for the $\Theta_\beta$, 
and finally for the $\Theta_\gamma$ we have 
\be
F_\gamma:
&\quad
y^3,y^5,y^7\in\left\{0,\frac12\right\},
\qquad
y^1\in\left\{\frac14,\frac34\right\},
\qquad
(y^2,y^4,y^6)\ \text{free}.
\label{eq:F-gamma}
\ee
Each connected covering-space component is locally
$T^3\times\frac{\mathbb{R}^4}{\mathbb{Z}_2}$
where the local $\mathbb{Z}_2$ acts by inversion on the four transverse
coordinates. We mentioned previously that globally, the other generators identify several of the sixteen
covering-space components.  For instance, four representatives of the
$F_\alpha$ orbits may be chosen as
\begin{equation}
(y^1,y^2,y^3,y^4)
=
(0,0,0,0),\quad
\left(0,0,\frac12,0\right),\quad
\left(0,0,0,\frac12\right),\quad
\left(0,0,\frac12,\frac12\right).
\end{equation}
Analyzing the locus of the orientifolds with respect to each involution we derive their intersections. 
These covering-space incidences are given in the table \ref{tab:incidences}. 
\begin{table}[ht!]\label{Incidences}
\centering
\begin{tabular}{@{}lll@{}}
\toprule
Singular locus & Covering-space orientifold loci & Local sectors \\
\midrule
$F_\alpha$ & ${\rm O6}_1$/${\rm O6}_2$ & one O6/O6 pair \\
$F_\beta$ & ${\rm O6}_1$/${\rm O6}_3$ & one ${\rm O6}/{\rm O6}$ pair \\
$F_\gamma$ & ${\rm O2}$/${\rm O6}_2$ , ${\rm O6}_3$/${\rm O6}_4$
& one O2/O6 sector and one O6/O6 sector \\
\bottomrule
\end{tabular}
\caption{Incidences before resolving the local
$\mathbb{C}^2/\mathbb{Z}_2$ singularities.}
\label{tab:incidences}
\end{table}
\footnote{Note that the apparent fourfold incidence at $F_\gamma$ is separated along the longitudinal coordinate $y^2$.}

Now we go back to what the resolutions imply for the local charts near the fixed points and their mappings to the global tori coordinates 
for a better understanding of the orientifold loci on the resolved manifold. 
Let $(z_1,z_2)$ be complex coordinates on the four-dimensional transverse space such that the local orbifold generator acts as
\begin{equation}
\mathbb{Z}_2: (z_1,z_2)\longmapsto(-z_1,-z_2).
\end{equation}
The quotient $\mathbb{C}^2/\mathbb{Z}_2$ is described by the invariant coordinates 
\begin{equation}
U=z_1^2,
\qquad
V=z_2^2,
\qquad
W=z_1z_2,
\end{equation}
which obey
\begin{equation}
UV-W^2=0.
\label{eq:A1-cone}
\end{equation}
Equation~\eqref{eq:A1-cone} is the $A_1$ surface singularity as a hypersurface embedded in $\mathbb{C}^3$.
The singularity in the invariant coordinates is at 
\begin{equation}
    U=V=W=0 \,. 
\end{equation}
The space of fixed direction approaching this singularity is described by a vector modulo the scalings. 
We define the vector as 
\begin{equation}
    [p:q]\rightarrow [\lambda p:\lambda q] \, \text{for all} \,\lambda \equiv \mathbb{R}  \,. 
\end{equation}
Here, $[p : q]$ denotes the ratio of two vectors.
We see that 
\begin{equation}
    [p:q] \equiv \mathbb{CP}^1 
\end{equation}
define the exceptional divisor $E$ of the resolved space. This is precisely the bolt sphere of the Eguchi--Hanson space.

We can parametrize the affine chart on the exceptional sphere. 
Consider for $p \neq 0$ 
\begin{equation}
    [p:q]\rightarrow [1:\frac{q}{p}] \,, 
\end{equation}
with one complex parameter
\begin{equation}
    \omega = \frac{q}{p} \,. 
\end{equation}
It covers the whole sphere but omits the direction $[0:1]$, which is $\omega \to \infty$. \footnote{It is clear that this is merely an artifact of 
choosing a particular chart, these direction can be considered by looking at another chart parameterized by $\tilde{
\omega}= \frac{p}{q}$ when $q\neq 0$.} 
For this chart we define
\begin{equation}
t=z_1^2,
\qquad
w=\frac{z_2}{z_1}.
\label{eq:t-w}
\end{equation}
In terms of invariant coordinates it is
\begin{equation}
U=t,
\qquad
W=tw,
\qquad
V=tw^2.
\label{eq:resolution-map}
\end{equation}

We see that this is a map from the resolved space to that of the singular space. The exceptional divisor E, 
is a map from a complex space defined as $E= (t=0, w)$  to the singular point. 
Clearly then we see that $t$ parametrizes the points orthogonal to E and $\omega$ parametrizes those tangent.

To make things more transparent we show that in terms of the metric of the Eguchi--Hanson space in Appendix \ref{Appendix Eguchi Hanson}, 
the near-bolt metric form can be mapped to coordinates on the affine chart $w$ and normal coordinate $t$ by 
\begin{equation} \label{singular patch to EH}
    t= \rho e^{i\chi} ,\quad w = e^{i\phi}\tan{\frac{\theta}{2}} \,. 
\end{equation}
Clearly, $t=0$ is mapped to the $r=\rho_0 $ (or bolt) of the Eguchi--Hanson space with $\chi$ parameterizing the collapsing normal angular coordinates to the bolt. 

We now first consider $F_\alpha$ near the representative
$(y^1,y^2,y^3,y^4)=(0,0,0,0)$.  Let $x^i$ denote the corresponding local
coordinates and define
\begin{equation}
z_1=x^1+i x^2,
\qquad
z_2=x^3+i x^4.
\end{equation}
The locus of the two orientifolds in table \ref{tab:incidences} on covering-space branches are
\begin{align}
{\rm O6}^{(1)} :L_{\mathrm{I}}
&:
\operatorname{Re}(z_1)=\operatorname{Re}(z_2)=0,
\label{eq:LI}
\\
{\rm O6}^{(2)}:L_{\mathrm{R}}
&:
\operatorname{Im}(z_1)=\operatorname{Im}(z_2)=0.
\label{eq:LR}
\end{align}
Considering the chart in \eqref{eq:resolution-map}, write
\begin{equation}
z_1=a_1+i b_1,
\qquad
z_2=a_2+i b_2.
\end{equation}
On the $L_{\mathrm{R}}$ locus one has
\begin{equation}
t=a_1^2\geq0,
\qquad
w=\frac{a_2}{a_1}\in\mathbb{R},
\end{equation}
whereas on $L_{\mathrm{I}}$ one has
\begin{equation}
t=-b_1^2\leq0,
\qquad
w=\frac{b_2}{b_1}\in\mathbb{R}.
\end{equation}
Hence the locus of the two planes on the singular patch are two half-surfaces
\begin{align}
\tilde L_{\mathrm{R}}
&:
t\geq0,
\qquad w\in\mathbb{R},
\\
\tilde L_{\mathrm{I}}
&:
t\leq0,
\qquad w\in\mathbb{R}.
\end{align}
Both meet the exceptional divisor at
\begin{equation}
E_{\mathbb{R}}
=
\left\{t=0,\ w\in\mathbb{R}\cup\{\infty\}\right\}
\cong\mathbb{RP}^1
\cong S^1 \,, 
\label{eq:real-exceptional-locus}
\end{equation}
which is a smooth non-self-intersecting surface on the $E$. 
Physically it describes that the two O6 planes sit on a great circle on $S^2$. 
This can be seen by noticing that in the orthogonal directions to the exceptional sphere the two orientifolds are clearly in different non-intersecting regimes parametrized by $t$. On the exceptional sphere, where $ \omega$ parametrizes a complex plane, we have the planes sitting on a common locus with $\omega\in
 \mathbb{R}$ which are circles $\phi= 0,\pi$ from \eqref{singular patch to EH}. 
 
 We have shown that the orientifolds, once the orbifolds points are resolved, do not intersect each other but instead sit on top of each other, 
 as after all, on a smooth space that manifests the involution symmetry the two O6 planes are in essentially the one and same plane. 
One must keep in mind that the choice of affine chart excludes a particular trajectory corresponding to $z_1=0$ axis but that is well covered in another well defined chart with $z_1$ and $z_2$ interchanged.

The same analysis applies to the ${\rm O6}_1/{\rm O6}_3$ branches at $F_\beta$ after a relabeling of the four transverse coordinates. 
It also applies to the ${\rm O6}_3/{\rm O6}_4$ sector of $F_\gamma$ at $y^2\in\{1/4,3/4\}$. 

However the conclusion for the ${\rm O2}/{\rm O6}_{4}$ intersection at the $F_\gamma$ fixed points needs more careful treatment. 
O2 planes sit precisely at the points of the blow-up and ${\rm O6}_{4}$ planes cover the whole Eguchi--Hanson space. 
Once we resolve the singularities of the space it is reasonable to expect that the O2-planes at those singularities cease to exist. 
This process can be mistaken to reduce the RR charge of the internal manifold consequently ruining the delicate balance of no net charge inside it simply via tadpole cancellation conditions. This seems to suggest that resolution of the geometry could potentially change the tadpole cancellations and puts the consistency of the solution at risk.

However we cure this pathology in the next section with a more careful analysis of the Bianchi identities and the charges of the O-planes.

\subsection{First Pontryagin class contributions} 

In this section we show that the suggestive pathology presented in the last subsection can be cured once we take into consideration the charges induced by curvature couplings of the Wess--Zumino terms in the orientifold plane action. This is where the role of ${\rm O6}_{(4)}$ wrapping the Eguchi--Hanson space becomes relevant.
We now show precisely that these planes make up for the missing O2 charge in the manifold. 
To demonstrate this, let us examine the higher-derivative contributions to the Chern–Simons term in the orientifold action
\be
S_{\rm CS/O6} = - 2 \mu_6 \int_7 \left[ C_7 + C_3 \frac{(4 \pi^2 \alpha')^2}{96} p_1(R_{\rm Tangent}) \right] \,, 
\ee
where we have ignored further higher order terms. 
The $\mu_6$ is the D6 charge/tension, 
given by setting $p=6$ in (restoring momentarily the string length) 
\be
\mu_p = \frac{(\alpha')^{-(p+1)/2}}{(2 \pi)^p} \,. 
\ee 
Notice crucially that 
\be
\mu_6 \times (4 \pi^2 \alpha')^2 
= \frac{(\alpha')^{-(6+1)/2}}{(2 \pi)^4} \times (2 \pi \alpha')^2 
= \frac{(\alpha')^{-(2+1)/2}}{(2 \pi)^2} 
= \mu_2 \,. 
\ee
The DBI piece is similarly 
\be
S_{\rm DBI/O6} = - 2 \mu_6 \int_7 e^{-\phi} \sqrt{-\gamma_7} + N_1 \times \int_7 e^{-\phi} \sqrt{-\gamma_3}  p_1(R_{\rm Tangent})  \,, 
\ee
where the coefficient of the $p_1$ terms, $N_1$, 
can be determined from supersymmetry arguments \cite{Bachas:1999um,Giddings:2001yu}. 
The DBI piece also contains higher-order terms that we do not need here. 
It is a standard result that the contribution of a single Eguchi--Hanson patch is\footnote{One way to see this is by taking into account that for the K3 we have $p_1(R_{\rm K3}) = 48$ which has 16 Eguchi--Hanson blow-ups.} 
\be
p_1(R_{\rm EH}) = -\frac12 \frac{1}{(2 \pi)^2} \int {\rm tr} R^2 = 3 \,. 
\ee
Now each O6 plane wrapping such a cycle would contribute to the total O2 charge since we have 
\begin{equation}\label{induced O2 charge}
S_{\rm CS/O6} = - 2 \mu_6 \int_7 \left[ C_7 + C_3 \frac{(4 \pi^2 \alpha')^2}{96} \times 3 \right]
= - 2 \mu_6 \int_7 C_7 
- \frac{1}{16} \mu_2 \int_3 C_3  \,.  
\end{equation}

Looking at the second term in the above expression makes it clear that one O6-plane induces a charge equivalent to half that of an O2-plane on flat space. 
This is where we recall the discussion in the paragraph below \eqref{O2 tadpole}, where we showed that for the $O_2$ planes sitting on the $\mathbb{Z}_2$ invariant locus their charge gets reduced to half of their charge on a flat space. 
In our construction we have $2^3$ O6-planes wrapping each of the three harmonic 4-forms $\psi_1$, $\psi_2$, $\psi_3$ and $\psi_4$, 
making in total $4 \times 2^3 = 32$. 
This happens because these orientifolds are localized in the Poincare-dual 3-cycles: $\phi_1$, $\phi_2$, $\phi_3$ and $\phi_4$. 
On top of that on each four-cycle wrapped by each of the O6-planes there are 4 Eguchi--Hanson blow-ups with each having $p_1=3$. 
Considering the exact number of O6 planes wrapping the blowup cycles which are 32 the total induced charge is $32\times \frac{1}{16}$ $=2$, which matches exactly the prediction in \eqref{O2 tadpole}.

The CS term also determines the effective contribution due to the higher order terms in the DBI piece. 
Since the O6-planes are mutually supersymmetric, the induced tension for the effective O2 object has to be mutually supersymmetric with the O6-planes. 
Therefore the tension has to be the tension of a standard O2, that is, negative. 
This means that the higher curvature terms in the DBI piece unavoidably conspire to produce a DBI piece of the form 
\be
S_{\rm DBI,eff \, O2} = - 8 \mu_2 \int_3 e^{-\phi} \sqrt{-\gamma_3} \,, 
\ee
which essentially determines $N_1$. 
To summarize we have an induced O2 effect from the O6-planes that behaves as 
\be
S_{\rm eff \, O2} = - 8 \mu_2 \int_3 e^{-\phi} \sqrt{-\gamma_3} - 8 \mu_2 \int_3 C_3 \,, 
\ee
where $\mu_2$ is the magnitude of the D2 tension/charge. 
In this analysis we remain relatively unbothered about D6-branes and D2-brane charges because they are free to move in the internal space and not necessarily sit on the points which are resolved. We come here to a rather elegant conclusion that for a suitable orbifold with a possible resolution the tadpole cancellation conditions remain invariant under the resolution of the geometry. The solution remains insensitive to the microscopic details of fixed points. This physically makes sense as the singular configuration can be achieved by just taking the radius of the exceptional sphere to zero smoothly.

\subsection{Counting the active ingredients}\label{Counting the active ingredients}

In the previous section we showed the configuration for Op planes in the internal manifold after the resolution. We also showed how the moduli associated with the resolution of fixed points, discussed in Sec. 3.1, are stabilized in Sec. 3.2, However as stated, we only resorted to the heuristics of the stabilization rather than precise numerical calculation. The main reason for this restriction was that we had not yet distinguished between the would-be local blow-up modes of the orbifold singularity and the physical blow-up moduli which are globally defined and consistent with the orientifold projection.
Distinguishing between these two is difficult from the perspective of the covering-torus picture, as it is agnostic to the details of the blow-ups. Only after understanding how the orientifolds lift to the blow-up patches can the parities of the exceptional cycles and forms be determined unambiguously. 
Here we now analyze the orientifold parities of the exceptional cycles and of the dual four-forms supporting the $F_4$ fluxes used in their stabilization.

Firstly we must note that the parity of the cycles with respect to one of the orientifold planes determines its parity with respect to all representatives. 
This can be seen from the simple following argument. 
Let $g\in G$ denote an element of the orbifold group, with \begin{equation}
X_7=\frac{\mathbb T^7}{G}, \end{equation} 
and let $\sigma$ denote the covering-space representative of the basic orientifold action. A physical global form $\omega$ is $G$-invariant and therefore satisfies 
\begin{equation} 
g^*\omega=\omega, \qquad \sigma^*\omega=\epsilon_\omega\omega, \qquad \epsilon_\omega=\pm1 \,. 
\end{equation}
The other covering-space representatives of the orientifold action are of the form $\sigma g$. Their action on $\omega$ is
\begin{equation} 
(\sigma g)^*\omega = g^*\sigma^*\omega = \epsilon_\omega g^*\omega = \epsilon_\omega\omega. 
\end{equation}
Thus all orientifold representatives assign the same parity to a global $G$-invariant form. The qualification that the form must first be globally defined is essential. An individual exceptional sphere or differential form in one covering patch is generally not by itself an object on the quotient. To compute its physical parity one must first return to
the chosen patch with an orbifold element.

Let us begin with the globally defined bulk moduli $s^i$ which parametrize the deformation of the structure form basis for the G2  in \eqref{3basis}. Each of the basis elements $\phi_i$ constituting $\Phi$ are invariant under the orbifold involution, and hence these are globally well defined. However, they are parity odd with respect to orientifold planes as can be seen from the table \ref{tab:branes}. 
Mathematically, consider the O2 action on the G2 structure  
\begin{equation}
\sigma^*\Phi=-\Phi, \qquad \sigma^*\Psi=+\Psi  \,. 
\end{equation}
It follows that the metric moduli multiply elements of the odd third cohomology 
\begin{equation}
[\phi_i]\in H^3_-(X_7) \,. 
\end{equation}
Their Poincar\'e-dual four-forms belong to $H^4_+(X_7)$. The opposite parities of a three-form and its Poincar\'e-dual four-form follow from the fact that $\sigma$ reverses the orientation of the seven-dimensional manifold.

We now apply the same analysis to the exceptional cycles introduced in eq.~\eqref{New periods} 
and to their corresponding forms introduced in eq.~\eqref{exceptional forms}. 
The expression 
\begin{equation}
\chi_{(\tau,n,a)} \simeq E_{\tau,n}\times S^1_a 
\end{equation}
denotes the local exceptional three-cycle, 
where $E_{\tau,n}\simeq S^2$ is the exceptional sphere. The associated localized three-form is 
\begin{equation} 
\zeta_{(\tau,n,a)} \simeq \nu_{\tau,n}\wedge d\xi_\tau^a, \end{equation}
where $\nu_{\tau,n}$ is the localized harmonic two-form on the Eguchi--Hanson space and $\xi_\tau^a$, $a=1,2,3$, are coordinates along the corresponding singular three-torus. We denote the Poincar\'e-dual four-forms by $\tilde\zeta^{(\tau,n,a)}$. 
Its Poincar\'e-dual four-form has the opposite sign because the ambient seven-manifold orientation is reversed. 
For an oriented cycle, the three-cycle $\chi_{(\tau,n,a)}$, it has the same parity as the three-form whose period detects it. Its Poincar\'e-dual four-cycle has the opposite parity because the ambient seven-dimensional orientation is reversed.
We are therefore interested in the pairs 
\begin{equation} 
\chi_{(\tau,n,a)}\in H_{3}^{(-)}(X_7) \ , \qquad \tilde\chi^{(\tau,n,a)}\in H_{4}^{(+)}(X_7) \ . 
\end{equation}

For an $\alpha$-type component, the relevant orientifold lift acts on the exceptional sphere by a conjugation-type involution whose fixed locus is $\mathbb{RP}^1\subset\mathbb{CP}^1$. It therefore reverses the orientation of the exceptional sphere. This can be seen explicitly by looking at the local patch defined by coordinates $(z_1,z_2)$. The orientifold representative $O_6^{(1)},O_6^{(2)}$ which fixes the patch acts as 
\begin{equation}
    (z_1,z_2) \rightarrow (\bar{z}_1,\bar{z}_2) \,. 
\end{equation}
On the exceptional cycle coordinate parametrized by $w$, 
in one affine chart it acts as (from eq. ~\eqref{eq:t-w}) 
\begin{equation}
    w \longmapsto \bar{w} \,, 
\end{equation}
which is clearly an orientation-reversing operation on $E_{\alpha,n}\simeq\mathbb{CP}^1$, 
consequently 
\begin{equation} \sigma^*\nu_{\alpha n}=-\nu_{\alpha n}. 
\end{equation} 
Using the tangent-coordinate ordering $
\xi_\alpha=(y^7,y^6,y^5) $
the corresponding one-forms transform as 
\begin{equation} 
\sigma^* \left(dy^7,dy^6,dy^5\right) = \left(+dy^7,-dy^6,+dy^5\right) \,. 
\end{equation}
Thus each $\alpha$-type component retains 
\begin{equation} \chi_{(\alpha,n,1)}, \ \chi_{(\alpha,n,3)} \in H_{3}^{(-)}(X_7), \qquad \tilde\chi^{(\alpha,n,1)}, \ \tilde\chi^{(\alpha,n,3)} \in H_{4}^{(+)}(X_7). \end{equation}
The remaining form $\chi_{\alpha,n,2}$ is a genuine global class in $H_{3(X_7)}^{(+)}$, but it does not describe an orientifold-compatible metric deformation.
Similarly for $\beta$ type fixed points the surviving classes are \begin{equation}
\chi_{(\beta,n,1)} \, , \ \chi_{(\beta,n,2)} \in H_{3}^{(-)}(X_7) \ , \qquad \tilde\chi^{(\beta,n,1)} \, , \ \tilde\chi^{(\beta,n,2)} \in H_{4}^{(+)}(X_7) \ . 
\end{equation}
The third form $\chi_{(\beta,n,3)}$ belongs to $H_{3(X_7)}^{(+)}$ and is projected out of the orientifold metric sector. 

The $\gamma$-type fixed components require an additional step. Let $P_{\rm O2/O6}$ denote the representative associated with the O2/O6 sector. 
Its action on the resolved transverse space descends to the identity on the exceptional sphere, so that 
\begin{equation} 
P_{\rm O2/O6}^*\nu_{\gamma,n} = +\nu_{\gamma,n} \,. 
\end{equation}
However, the representative associated with the ${\rm O6}_3/{\rm O6}_4$ sector, denoted by $P_{\rm O6/O6}$, acts by conjugation on the exceptional sphere and hence 
\begin{equation} 
P_{\rm O6/O6}^*\nu_{\gamma,n} = -\nu_{\gamma,n} \,. 
\end{equation} 
This apparent disagreement signals that $\nu_{\gamma,n}$ does not itself define a global two-form on the resolved quotient.
However, the candidate global exceptional three-forms are obtained by combining $\nu_{\gamma}$ with tangent one-forms, defined as 
\begin{equation}
    \zeta_1=\nu_{\gamma} \wedge d y^2, \quad \zeta_2=\nu_{\gamma} \wedge d y^4, \quad \zeta_3=\nu_{\gamma} \wedge d y^6 .
\end{equation}
Therefore 
\begin{equation}
    P_{\rm O2/O6}^*\left(\chi_1, \chi_2, \chi_3\right)=\left(-\chi_1,-\chi_2,-\chi_3\right) \,. 
\end{equation}
For the $P_{\rm O6/O6}$ representatives 
\begin{equation}
    P_{\rm O6/O6}^* \nu_\gamma=-\nu_\gamma \,, 
\end{equation}
consequently 
\begin{equation}
     P_{\rm O6/O6}^* \left(\chi_1, \chi_2, \chi_3\right)=\left(+\chi_1,-\chi_2,-\chi_3\right) \,. 
\end{equation}
The two orientifold representatives agree on $\chi_{\gamma,n,2}$ and $\chi_{\gamma,n,3}$ but disagree on $\chi_{\gamma,n,1}$. Since different representatives must assign the same parity to a global orbifold-invariant form, this disagreement shows that $\chi_{\gamma,n,1}$ does not descend to the resolved quotient. The physical globally defined classes are therefore 
\begin{equation}
\chi_{(\gamma,n,2)} \, , \ \chi_{(\gamma,n,3)} \in H_{3}^{(-)}(X_7)\ , \qquad \tilde\chi^{(\gamma,n,2)} \, , \ \tilde\chi^{(\gamma,n,3)} \in H_{4}^{(+)}(X_7) \ . 
\end{equation}
This conclusion can be restated in terms of the residual orbifold involution as 
\begin{equation}
\kappa \equiv \Theta_\alpha \Theta_\beta, \quad \kappa:\left(y^2, y^4, y^6\right) \mapsto\left(y^2+\frac{1}{2},-y^4,-y^6\right) \,. 
\end{equation}
The local resolved neighborhood is therefore 
\begin{equation} 
\frac{ \left(\mathbb R^4/\mathbb Z_2\right)\times T^3_\gamma }{ \langle\kappa\rangle },
\end{equation}
and for locally defined forms to descend to global forms, they must be $\kappa$-invariant. 
One observes that the two orientifold representatives obey 
\begin{equation} 
P_{\rm O6/O6} = P_{\rm O2/O6}\kappa  \,, 
\end{equation} 
and their action is consistent once we choose 
\begin{equation}
\kappa^*\nu_{\gamma }=-\nu_{\gamma } \,. 
\end{equation}
We have therefore selected the second of the two topologically distinct $\gamma$-resolutions discussed in \cite{Acharya:1996fx}.
The three local candidates obey 
\begin{equation} 
\kappa^* \left( \chi_{\gamma 1}, \chi_{\gamma 2}, \chi_{\gamma 3} \right) = \left( -\chi_{\gamma 1}, +\chi_{\gamma 2}, +\chi_{\gamma 3} \right) \,, 
\end{equation} 
further confirming the blow-up moduli correspond to the global classes $\chi_{\gamma2}, \chi_{\gamma 3}$.

We can now determine the Betti numbers. 
There are four inequivalent $\alpha$-type components and four inequivalent $\beta$-type components, 
while each contributes 
\begin{equation} 
(\Delta b_2,\Delta b_3)=(1,3) \, . 
\end{equation}
Together, these eight components contribute 
\begin{equation} 
(\Delta b_2,\Delta b_3)_{\alpha+\beta} = (8,24) \, . 
\end{equation}
Turning now to the $\gamma$ involution, 
there are eight inequivalent $\gamma$-type components. 
For the resolution selected above, the complete $\gamma$ sector therefore contributes 
\begin{equation}
(\Delta b_2,\Delta b_3)_{\gamma} = (0,16) \,. 
\end{equation}
Including eventually the seven untwisted three-forms gives the total counting 
\begin{equation} 
b_2  = 8 \quad ,\quad b_3 = 7+24+16 = 47 \ . 
\end{equation}
This agrees with the general result of \cite{Acharya:1996fx,Joyce2} 
\be 
b_2=8+l \ , \qquad b_3=47-l \ , 
\ee 
where $l$ is the number of $\gamma$-components resolved using the first choice, 
which contributes to the counting $(\Delta b_2,\Delta b_3)=(1,1)$. 
Our resolution uses the second choice at all eight $\gamma$-components and therefore corresponds to $l=0$. 
After imposing the orientifold projection, the second and third cohomology split as 
\begin{equation} 
b_2^-=8 \ , \qquad b_2^+=0 \ ,
\end{equation} 
and 
\begin{equation} 
b_3^- = 7_{\mathrm{bulk}} + 4(2)_\alpha + 4(2)_\beta + 8(2)_\gamma = 39  \ , \quad 
b_3^+ = 4(1)_\alpha + 4(1)_\beta = 8 \,. 
\end{equation} 
Since $\sigma$ reverses the orientation of $X_7$, 
Poincar\'e duality gives \begin{equation} 
b_4^+=b_3^-=39, \qquad b_4^-=b_3^+=8. \end{equation}
For clarity, the contribution and parity content of each type of resolved component are summarized in Table ~\ref{tab:exceptional-cycle-parities}. 
\begin{table}[t] \centering \small \renewcommand{\arraystretch}{1.4} \begin{tabular}{|c|c|c|c|c|c|} \hline sector & $N_\tau$ & $(\Delta b_2,\Delta b_3)$ per component & global cycle--form pairs & odd pairs & even pairs \\ \hline $\alpha$ & $4$ & $(1,3)$ & $\chi_{1,2,3} \leftrightarrow \zeta_{1,2,3}$ & $\chi_{1,3} \leftrightarrow \zeta_{1,3}$ & $\chi_2 \leftrightarrow \zeta_2$ \\ $\beta$ & $4$ & $(1,3)$ & $\chi_{1,2,3} \leftrightarrow \zeta_{1,2,3}$ & $\chi_{1,2} \leftrightarrow \zeta_{1,2}$ & $\chi_3 \leftrightarrow \zeta_3$ \\ $\gamma$ & $8$ & $(0,2)$ & $\chi_{2,3} \leftrightarrow \zeta_{2,3}$ & $\chi_{2,3} \leftrightarrow \zeta_{2,3}$ & --- \\ \hline \end{tabular} \caption{ Global exceptional three-cycles and their associated three-form representatives. The fourth column lists the cycle--form pairs which exist on the resolved G2 manifold before the orientifold projection. The fifth and sixth columns give their decomposition into $H_{3,-}\leftrightarrow H^3_-$ and $H_{3,+}\leftrightarrow H^3_+$, respectively. For the chosen $\gamma$-resolution, the local pair $\chi_{\gamma,n,1}\leftrightarrow\zeta_{\gamma,n,1}$ is not $\kappa$-invariant and therefore does not define a global class. } \label{tab:exceptional-cycle-parities} 
\end{table}

Consequently, every surviving exceptional geometric modulus has an even dual four-form on which $F_4$ flux is allowed. This statement establishes that the orientifold parity does not obstruct the stabilization mechanism. Flux quantization and tadpole conditions may constrain the allowed integer values, but they do not remove the cohomological pairing between a surviving blow-up modulus and its dual four-form.
The projected-out components should be understood as fields absent from the consistent orientifold truncation, rather than as physical scalar fields which remain to be stabilized dynamically.

The analysis of Sec.~\ref{Bosonic reduction cross-check} must therefore be restricted to the parity-allowed subspace. 
For the $\alpha$-type involution, the Hodge matrix restricted to the surviving blow-up moduli and their corresponding fluxes has a block-diagonal form
\begin{equation} 
\begin{aligned} 
K_{\text{rel}} &= K_{\left(u_1,s^5,s^7\right)} \oplus K_{\left(u_3,s^1,s^6\right)} \oplus K_{\text{spectator}} +\mathcal O(u_au_b).  \end{aligned}
\end{equation}
The block involving $(u_2,s^3,s^4)$ is absent because $u_2$ and its dual exceptional flux coefficient are projected out.
The explicit numerical Hodge sub-block displayed above was derived for an $\alpha$-type neighbourhood in Sec.~\ref{Bosonic reduction cross-check}. Its application to the surviving $\alpha$ directions is unchanged. For the $\gamma$ sector, the same cohomological mechanism applies to the globally defined $\kappa$-invariant forms. We expect the scaling for the blow-up modes to remain the same.

\subsection{A specific example}\label{A specific example}

Now we study the stabilization of the blow-up moduli for the flux background defined in section \ref{Flux example with D6-branes and D2 scali}. Given the table ~\ref{tab:exceptional-cycle-parities}, we calculate the numerics for stabilization of the relevant moduli. 
Without loss of generality, we resolve one fixed component of the $\Theta_{\alpha}$ involution, here the cycles consistent with the orientifold parity are $\chi_1, \chi_3$. 
The bulk flux choice is summarized as 
\begin{equation}
 f^i=-f\,(1,1,1,1,1,1,1),
 \qquad
 h^i=h\,(1,1,1,1,-1,-1,-1),
 \qquad
 F_0=m,
\end{equation}
with $f,h,m>0$.

In the harmonic-form convention of eqs.~\eqref{3basis} and~\eqref{4basis}, the two $\Theta_{\alpha}$ blow-up periods are paired with the bulk periods according to
\begin{equation}
 (A_a,B_a)=(5,7),(1,6).
\end{equation}
The resolved Einstein-frame volume, to quadratic order in the exceptional periods, is
\begin{equation}
 \mathcal V
 =
 V_0\left(1-\frac{8}{3}\sum_{a=1}^{2}d_a u_a^2\right)
 +{\cal O}(u^4),
 \qquad
 V_0=\left(\prod_{i=1}^{7}s^i\right)^{1/3},
\end{equation}
where
\begin{equation}
 d_a=\frac{1}{s^{A_a}s^{B_a}} \,, 
\end{equation}
and the summation is over the two relevant moduli. 
In the physical Einstein-frame convention, the blow-up-dependent superpotential is expanded as
\begin{equation}
 P
 =
 P_0
 +\sum_{a=1}^{2}L_a u_a
 +\frac{1}{2}\sum_{a=1}^{2}X_a u_a^2
 +{\cal O}(u^3),
\end{equation}
where $P_0$ is the superpotential of the untwisted bulk fields. 
The exceptional-flux coefficient is
\begin{equation}
 L_a =  \frac{1}{8}e^{\phi/4}V_0^{-2}q_a \, , 
\end{equation}
and the quadratic coefficient is
\begin{equation}
 X_a
 =
 d_a\Bigg\{
 \frac{2}{3}m e^{5\phi/4}V_0^{-1}
 +e^{-\phi/2}V_0^{-1}
 \left[
 \frac{2}{3}\sum_{i=1}^{7}\frac{h^i}{s^i}
 +\frac{h^{A_a}}{s^{A_a}}
 +\frac{h^{B_a}}{s^{B_a}}
 \right]
 +\frac{4}{3}e^{\phi/4}V_0^{-2}
 \sum_{i=1}^{7}f^i s^i
 \Bigg\}.
\label{eq:Ba-D6-D2}
\end{equation}

The ingredients that are needed to evaluate \eqref{eq:Ba-D6-D2} related to the bulk fields can be imported directly from subsection \ref{Flux example with D6-branes and D2 scali}, taking in particular into account that $V_0$ is given in formula \eqref{Ef-volume} which is the volume without any blow-up moduli contributions. 
For the first exceptional period, $(A_1,B_1)=(5,7)$, 
therefore we get 
\begin{equation}
 X_1
 =
 -\frac{3584}{16875}
 \frac{h^3m^2}{f^4(s^5)^2}
 =-\frac{2^{105 / 8} 7^{1 / 8}}{3^{9 / 2} 5^{25 / 4}} \frac{h^{9 / 2} m^{25 / 8}}{f^{53 / 8}} \,. 
\end{equation}
For the third exceptional period, $(A_3,B_3)=(1,6)$, similarly
it follows that 
\begin{equation}
 X_3
 =
 -\frac{2048}{16875}
 \frac{h^3m^2}{f^4s^1s^6}=-\frac{2^{121 / 8} 7^{1 / 8}}{3^{9 / 2} 5^{25 / 4}} \frac{h^{9 / 2} m^{25 / 8}}{f^{53 / 8}} \,. 
\end{equation}
To our advantage all exceptional periods therefore receive nonvanishing quadratic coefficients.

Now we extremize the superpotential wrt the blow-up moduli and the leading supersymmetry equations are
\begin{equation}
 P_{u_a}=L_a+X_a u_a+ {\cal O} (u^2)=0,
\end{equation}
with solutions up to leading order given by 
\begin{equation}
 u_a^{\star}=-\frac{L_a}{X_a}.
\end{equation}
Using the explicit $L_a$ and $X_a$ gives
\begin{equation}
 \frac{u_1^{\star}}{\sqrt{s^5s^7}}
 =\frac{q_1}{16f} ,
 \qquad
 \frac{u_3^{\star}}{\sqrt{s^1s^6}}
 =\frac{\sqrt7\,q_3}{64f}.
\end{equation}

The relative correction to the Einstein-frame volume is therefore 
\begin{equation}
\frac{\Delta V}{V_0}
 =
 -\frac{1}{f^2}
 \left(
 \frac{q_1^2}{96}
 +\frac{7q_3^2}{1536}
 \right)
 + {\cal O} \left(\frac{q^4}{f^4}\right) \,. 
\end{equation}
Let us define for convenience 
\begin{equation}
Q_E^2\equiv112q_1^2+7q_3^2 \, , 
\end{equation}
for a normalized flux quanta on the exceptional cycles.
Thus the small-resolution expansion is controlled for
\begin{equation}
Q_E\ll f.
\end{equation}
We emphasize again that the untwisted solution is not an exact solution after turning on
$q_a\neq0$. Integrating out the blow-up modulus at leading order gives
\begin{equation}
\Delta P_\star
=
-\sum_{a=1}^2\frac{L_a^2}{2 X_a},
\end{equation}
and hence the shifts of the dilaton and untwisted geometric moduli begin
parametrically at
\begin{equation}
\frac{\delta s^i}{s^i} \sim \delta\phi \sim {\cal O} \left(\frac{Q_E^2}{f^2}\right) \, .  
\end{equation}
While stabilization of the blow-up cycle at a parametrically smaller scale than the bulk moduli is a desirable feature, one must carefully analyze whether the sizes of the blow-up cycles remain larger than the string length so that the supergravity approximation is still valid. 

To approximate that we compare the size of the exceptional sphere with the string length. 
Here we make use of relation of the radius of the seven torus with the bulk moduli and we get 
\begin{equation}
 R_1=R_3=R_6=s^{1/3},
 \qquad
 R_2=\frac{s^{1/3}}{\sqrt7},
 \qquad
 R_4=R_5=R_7=\sqrt7\,s^{1/3} \,. 
\end{equation}
Using
\begin{equation}
 u_a=-\frac{\pi}{2}R_a\rho_E^2n_a,
 \qquad
 \sum_an_a^2=1,
\end{equation}
for $a=1,3$ one finds
\begin{equation}
 \frac{\rho_E^2}{s^{2/3}}
 =
 \frac{Q_E}{32\pi f} \,. 
\end{equation}
The characteristic string-frame bulk radius is
\begin{equation}
 \frac{r_S^2}{\alpha'}=V_S^{2/7}
 =
 \left[
 \frac{125\,14^{1/4}\sqrt5}{196}
 \right]^{2/7}
 \sqrt{\frac{f}{m}}
 \simeq
 1.3363\sqrt{\frac{f}{m}} \,, 
\end{equation}
therefore 
\begin{equation}
 \frac{\rho_S^2}{\alpha'}
 =
 \frac{1}{32\pi}
 \left[
 \frac{125\,14^{1/4}\sqrt5}{196}
 \right]^{2/7}
 \frac{Q_E}{\sqrt{fm}}
 \simeq
 0.01329\frac{Q_E}{\sqrt{fm}} \, .
\end{equation}
A parametrically controlled resolved regime requires simultaneously $\rho_S\gg\ell_s$ and $\rho_S\ll r_S$. Up to the displayed numerical factors, this gives
\begin{equation}
 \sqrt{fm}\ll Q_E\ll f \,. 
\end{equation}
For fixed $m$, this window can be realized by taking $Q_E\sim f^p$ with $1/2<p<1$. In this regime the exceptional sphere is large in string units, the resolution remains parametrically small relative to the bulk, and the bulk-modulus shifts vanish as $f^{2p-2}$.
In this regime, the supergravity approximation remains valid, with all relevant length scales in the manifold larger than the string length and yet blow-up remains small enough to not affect the leading order solution.
We hence state that the above solution remains scale separated at parametrically large flux choice on a smooth non singular G2 manifold.

\section{The brane dual}\label{Section 5:The brane dual}

Can we reconstruct a brane-dual geometry for these vacua, along the lines of \cite{Apers:2025pon,Apers:2026lgi}? We proceed in two steps. First, we use flux backtracking to determine the background geometry probed by the D4-branes, removing the \(F_4\) fluxes that they source. The resulting flow spans a finite radial interval, bounded by a strongly coupled singularity at one end and a weakly coupled singularity at the other. By tracking when the removed \(F_4\) terms become dominant, we identify the strongly coupled endpoint as the D4-brane locus. The corresponding low-energy absorption probability vanishes, suggesting that the brane degrees of freedom decouple from the flux-backtracked bulk. We then restore the \(F_4\) fluxes and ask whether the D4-brane backreaction can be described by the usual harmonic-superposition picture. Unlike in the DGKT vacua and in the scale-separated AdS$_3$ vacua of type IIB \cite{Apers:2026lgi,Apers:2026ahu}, this does not lead to a full analytic interpolating geometry. The D4-brane charges are correlated by tadpole cancellation, while the individual D4-branes are Freed--Witten anomalous in the \(H_3\) background, so an independent-stack harmonic superposition is not expected to solve the coupled system. A flow to the supersymmetric AdS$_3$ vacuum can nevertheless be obtained numerically.

\subsection{Flux backtracking to find the geometry probed by the D4-branes}

For the specific flux configuration under consideration, the untwisted superpotential is
\begin{equation}
P
=
\frac{1}{8}
\left[
\frac{m e^{5\phi/4}}{\mathcal V}
+
\frac{e^{-\phi/2}}{\mathcal V}
\sum_{i=1}^{7}\frac{h_i}{s^i}
+
\frac{e^{\phi/4}}{\mathcal V^2}
\sum_{i=1}^{7} e_i s^i
\right],
\qquad
\mathcal V=\left(\prod_{i=1}^{7}s^i\right)^{1/3},
\label{eq:full_superpotential}
\end{equation}
where, for the example of interest,
\begin{equation}
h_i=(h,h,h,h,-h,-h,-h),
\qquad
e_i=-f.
\end{equation}

We would now like to determine the geometry that is probed by the D4-branes. Following the flux-backtracking prescription, we remove the flux sourced by the branes whose backreaction we ultimately wish to restore. Since the D4-branes source the \(F_4\)-fluxes, this amounts to setting \(e_i=0\). The residual superpotential is therefore
\begin{equation}
P_{\rm res}
=
\frac{1}{8}
\left[
\frac{m e^{5\phi/4}}{\mathcal V}
+
\frac{e^{-\phi/2}}{\mathcal V}
\sum_{i=1}^{7}\frac{h_i}{s^i}
\right].
\label{eq:backtracked_superpotential}
\end{equation}

The standard flat domain-wall ansatz is
\begin{equation}
ds_{3,E}^2
=
dr^2
+
e^{2A(r)}
\left(
-dt^2+dx^2
\right),
\label{eq:domain_wall_ansatz}
\end{equation}
with all scalar fields depending only on the radial coordinate $r$. 

Using the symmetry of the flux configuration, we set
\begin{equation}
s^{1,2,3,4}=a,
\qquad
s^{5,6,7}=b,
\end{equation}
and define
\begin{equation}
M\equiv \frac{m e^{5\phi/4}}{8\mathcal V},
\qquad
H_+\equiv \frac{h e^{-\phi/2}}{8\mathcal V\,a},
\qquad
H_-\equiv -\frac{h e^{-\phi/2}}{8\mathcal V\,b},
\label{eq:MHpm_definitions}
\end{equation}
with $H_+>0$ and $H_-<0$. The residual superpotential is then
\begin{equation}
P_{\rm res}=M+4H_+ +3H_-.
\end{equation}
The corresponding BPS equations reduce to
\begin{align}
M'&=-8M^2,
\\
H_+'&=-4H_+\left(5H_+ +3H_-\right),
\\
H_-'&=-16H_-\left(H_+ +H_-\right),
\\
A'&=2\left(M+4H_+ +3H_-\right),
\label{eq:compact_bps_system}
\end{align}
where a prime denotes differentiation with respect to \(r\).
\subsubsection{Finite interval}
We introduce\begin{equation}
z\equiv-\frac{H_-}{H_+}>0,
\end{equation}
in terms of which the solution to the $H$-system can be written as
\begin{equation}
H_+
=
C\,\frac{(1+z)^8}{z^5},
\qquad
H_-
=
-C\,\frac{(1+z)^8}{z^4},
\label{eq:Hpm_solution}
\end{equation}
where $C>0$ is an integration constant. The radial coordinate is determined by
\begin{equation}
\frac{dr}{dz}
=
\frac{z^4}{4C(1+z)^9}.
\label{eq:r_z}
\end{equation}
This implies that
\begin{equation}
    \Delta r =\frac{1}{4C}\int_{0}^{\infty}\frac{z^4}{(1+z)^9} dz=\frac{1}{1120C}.
\end{equation}
Thus \(z\in(0,\infty)\) maps to a finite coordinate interval \(r\in(0,r_\star)\), with \(r_\star=1/(1120C)\).
Hence, the brane dual will be realised on a \emph{finite interval}. This is a consequence of having $H_3$-fluxes of both signs contributing.
\subsubsection{Explicit background solution}
The flow equations do not have a scaling solution as in the original flux backtracking examples \cite{Apers:2025pon}. Still, it is possible to obtain an explicit solution.
The first BPS equation is easily integrated as
\begin{equation}
    M(r) = \frac{1}{8(r-r_M)} = \frac{1}{8r}
\end{equation}
choosing $r_M=0$. For the warp factor we find 
\begin{equation}
    e^A = e^{A_0}\, r^{1/4}\frac{z^2}{(1+z)^{7/2}}.
\end{equation}
The radial profile of the moduli can then be expressed as
\begin{align}
e^\phi
&=
\left(\frac{h}{8C}\right)^{7/16}
(m r)^{-5/8} 
\frac{z^2}{(1+z)^{7/2}},
\\[0.5em]
\mathcal{V}
&=
\left(\frac{h}{8C}\right)^{35/64}
(m r)^{7/32} 
\frac{z^{5/2}}{(1+z)^{35/8}},
\\[0.5em]
a
&=
\left(\frac{h}{8C}\right)^{15/64}
(m r)^{3/32} 
\frac{z^{3/2}}{(1+z)^{15/8}},
\\[0.5em]
b
&=
\left(\frac{h}{8C}\right)^{15/64}
(m r)^{3/32} 
\frac{z^{1/2}}{(1+z)^{15/8}}.
\end{align}
The 10-dimensional string frame background then takes the form
\begin{align}
ds_{10,S}^2
={}&
\left(\frac{8C}{h}\right)^{7/8}m^{-3/4}
\left[
r^{-3/4}\frac{(1+z)^7}{z^4}\,dr^2
+
e^{2A_0}r^{-1/4}
\left(-dt^2+dx^2\right)
\right]
\nonumber\\
&+
\left(\frac{h}{8C}\right)^{3/8}m^{-1/4}
\frac{r^{-1/4}}{(1+z)^3}
\Big[
z^2\big((dy^1)^2+(dy^3)^2+(dy^6)^2\big)
\nonumber\\
&\hspace{7em}
+z^3(dy^2)^2
+z\big((dy^4)^2+(dy^5)^2+(dy^7)^2\big)
\Big].
\end{align}

The prefactor of the non-compact directions is independent of \(z\) and decreases with increasing \(r\). The geometry therefore exhibits the blueshift behaviour relative to the string scale discussed in \cite{Apers:2026lgi,Bedroya:2025ltj}. In contrast to the examples considered there, however, the finite radial interval prevents this blueshift from becoming arbitrarily large.

\subsubsection{The strongly coupled endpoint: \texorpdfstring{$z\rightarrow 0$}{z→0}}
We zoom in on the left endpoint of the interval, where we take $z\rightarrow 0$ to coincide with $r\rightarrow 0$. In that case $r\approx z^5$ and the metric approaches, up to constant prefactors,
\begin{align}
    ds_{10,S}^2
={}&d\rho^2 + \rho^{-10/9} (-dt^2+dx^2) + \rho^{2/3} \big((dy^1)^2+(dy^3)^2+(dy^6)^2\big)\\
& + \rho^{14/9} (dy^2)^2
+\rho^{-2/9}\big((dy^4)^2+(dy^5)^2+(dy^7)^2\big),
\end{align}
and
\begin{align}
    e^\phi \sim \rho^{-1}
\end{align}
where we defined $\rho:= r^{9/40}$. Interestingly, this resembles the background metric of the DGKT vacua with the scaling for the non-compact coordinates and the dilaton being identical. This is a strongly coupled singularity.

\subsubsection{The weakly coupled endpoint: \texorpdfstring{$z\to\infty$}{z→∞}} 

We can similarly zoom in on the other endpoint of the finite radial interval, corresponding to \(z\rightarrow\infty\). Denoting the endpoint by
\begin{equation}
r_{\star}=\frac{1}{1120C},
\end{equation}
and defining
\begin{equation}
\delta r:=r_{\star}-r,
\end{equation}
we have
\begin{equation}
\delta r
=
\frac{1}{4C}
\int_{z}^{\infty}
\frac{u^4}{(1+u)^9}\,du
\simeq
\frac{1}{16Cz^4}.
\end{equation}
Therefore,
\begin{equation}
\delta r\sim z^{-4}.
\end{equation}

We introduce a proper radial coordinate satisfying
\begin{equation}
\tilde{\rho}\sim(\delta r)^{5/8},
\end{equation}
in terms of which the ten-dimensional string-frame metric, up to constant prefactors, is given by
\begin{align}
ds_{10,S}^2
={}&d\tilde{\rho}^2 + (-dt^2+dx^2) + \tilde{\rho}^{2/5} \big((dy^1)^2+(dy^3)^2+(dy^6)^2\big)\\
& + (dy^2)^2
+\tilde{\rho}^{4/5}\big((dy^4)^2+(dy^5)^2+(dy^7)^2\big),
\end{align}
and
\begin{align}
e^\phi \sim \tilde{\rho}^{3/5}.
\end{align}
This endpoint is qualitatively different from the \(z\rightarrow0\) endpoint. In particular, the string coupling vanishes as \(\tilde{\rho}\rightarrow0\), so this is a weakly coupled singularity. 

\begin{figure}
    \centering
    \includegraphics[width=1\linewidth]{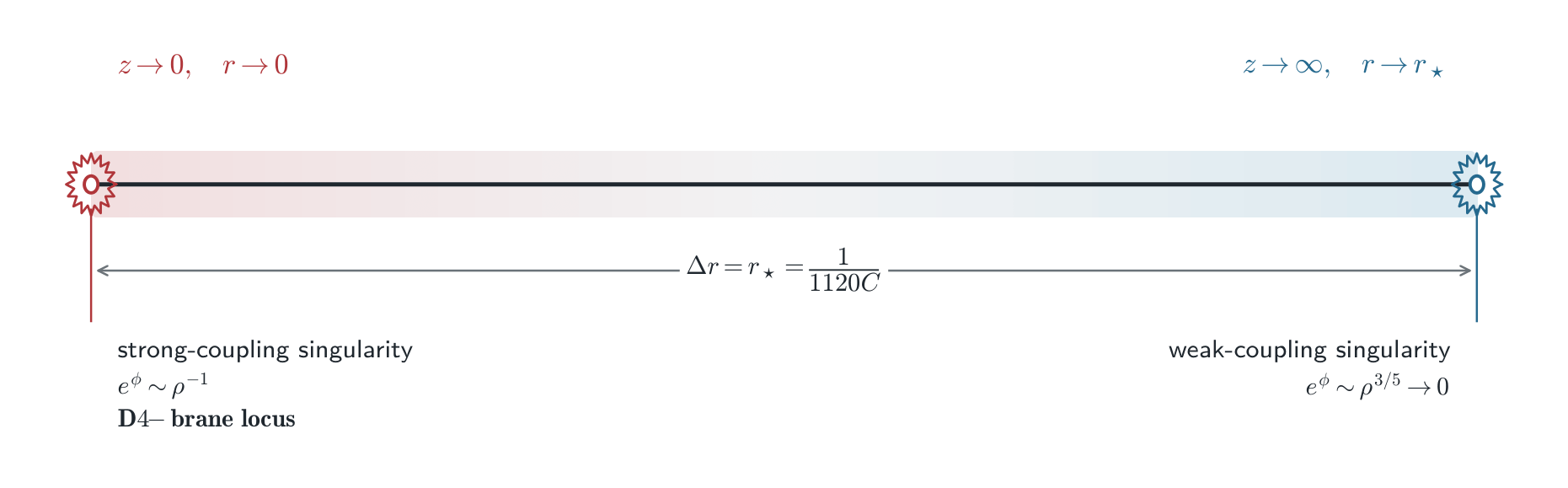}
    \caption{Applying flux-backtracking, we find an interval with a strongly coupled singularity on one end and a weakly coupled one on the other end.}
    \label{fig:interval}
\end{figure}

\subsubsection{Identifying the D4-brane locus}
With two singular endpoints, there are in principle two possible loci for the stack of D4-branes sourcing the $F_4$-fluxes. We identify the D4-brane locus by requiring that the flux terms sourced by the D4-branes become parametrically important near the corresponding endpoint.

We define
\begin{equation}
F_+
\equiv
-\frac{f e^{\phi/4}}{8\mathcal V^2}a,
\qquad
F_-
\equiv
-\frac{f e^{\phi/4}}{8\mathcal V^2}b,
\end{equation}
so that
\begin{equation}
P = M+4H_+ +3H_-+4F_+ +3F_-.
\end{equation}

Near the left endpoint, $z\rightarrow0$, we have
\begin{equation}
r\sim\rho^{40/9},
\qquad
z\sim\rho^{8/9},
\end{equation}
and therefore
\begin{equation}
M\sim\rho^{-40/9},
\qquad
H_+\sim\rho^{-40/9},
\qquad
H_-\sim \rho^{-32/9},
\end{equation}
while the removed $F_4$-terms scale as
\begin{equation}
F_+\sim \rho^{-44/9},
\qquad
F_-\sim \rho^{-52/9}.
\end{equation}
Thus, as $\rho\to0$, the $F_4$-dependent terms diverge more strongly than the other flux contributions. Equivalently, their relative importance decreases more rapidly as one moves away from the endpoint. This is precisely the behavior expected near a localized D4-brane source.

Near the right endpoint, on the other hand, we find the following
\begin{equation}
\delta r \sim \tilde{\rho}^{8/5},
\qquad
z \sim \tilde{\rho}^{-2/5}.
\end{equation}
The contributions to the superpotential then scale like
\begin{equation}
M \sim \tilde{\rho}^0,
\qquad
H_+ \sim z^3 \sim \tilde{\rho}^{-6/5},
\qquad
H_- \sim z^4 \sim \tilde{\rho}^{-8/5},
\end{equation}
and 
\begin{equation}
F_+ \sim z^3 \sim \tilde{\rho}^{-6/5},
\qquad
F_- \sim z^2 \sim \tilde{\rho}^{-4/5}.
\end{equation}

As the right endpoint is approached, the $F_4$-dependent terms therefore become parametrically suppressed relative to the $H_{-}$ contribution. 

We therefore identify the $z\to0$ endpoint as the D4-brane locus.

\subsubsection{Decoupling of the D4-branes}

Having identified the left endpoint of the interval as the locus of the
D4-brane stack, this is the region in which the flux-backtracked geometry
has to be glued to the $\mathrm{AdS}_3$ throat sourced by the branes.
We can therefore use the general analysis of \cite{Apers:2026ahu} 
to determine whether the degrees of freedom on the D4-branes decouple
from the bulk in the low-energy limit. In a one-modulus asymptotic
regime, the relevant information is the dimension of the AdS throat
together with the steepness $\lambda$ of the scalar potential,
\begin{equation}
    V_{\rm res}(\varphi)
    \sim
    -V_0 e^{-\lambda \varphi},
\end{equation}
where $\varphi$ denotes the canonically normalized modulus along the flux-backtracking trajectory.

Near the left endpoint,
\begin{equation}
    r\sim z^5,
    \qquad
    e^A\sim r^{1/4}z^2\sim r^{13/20},
\end{equation}
Thus, the asymptotic domain-wall metric is of the form
\begin{equation}
    ds_3^2
    =
    dr^2+r^{2q}ds_{1,1}^2,
    \qquad
    q=\frac{13}{20}.
\end{equation}
For an exponential potential in $d$ dimensions, the corresponding
scaling solution satisfies
\begin{equation}
    q=\frac{4}{(d-2)\lambda^2}.
\end{equation}
Since in the present case $d=3$, we obtain
\begin{equation}
    \lambda^2=\frac{4}{q}
    =\frac{80}{13}.
\end{equation}

We can now directly apply the absorption result of \cite{Apers:2026ahu}. 
The low-energy absorption probability therefore behaves as 
\begin{equation}
    \mathcal P_{\rm abs}
    \sim \omega^{\Delta_{\rm abs}},
    \qquad
    \Delta_{\rm abs}
    =
    \frac{\lambda^2}{\lambda^2-4}
    =
    \frac{20}{7}.
\end{equation}
Hence
\begin{equation}
    \mathcal P_{\rm abs}\sim\omega^{20/7}
    \longrightarrow 0,
    \qquad \omega\to0.
\end{equation}

The absorption probability therefore vanishes as $\omega\to0$.
Perturbations localized in the $\mathrm{AdS}_3$ throat cannot escape
into the flux-backtracked bulk at parametrically low energies, suggesting
that the D4-brane degrees of freedom decouple from the bulk.

\subsection{Placing D4-branes on top of the background}

We now ask what happens when the $F_4$ fluxes are restored on the flux-backtracked background of the previous subsection. The corresponding D4-branes wrap the three-cycles dual to the seven $F_4$ components and form domain walls in three dimensions.

An important difference from the simplest brane constructions is that these D4-branes cannot be varied independently. Their charges are correlated by the tadpole conditions, which retain information about the compactness of the internal space. For example, from (2.13) and (2.14), one may shift the flux while keeping the D2 tadpole unchanged according to
\begin{equation}
    F_4 \rightarrow F_4 + f'(\psi_1+\psi_5).
\end{equation}

This possibly pairs up the D4$_1$ and the D4$_5$ into a single effective brane, in the sense that it is impossible to change $F_4$ only in the $\psi_1$ direction; the compactness of the internal space automatically enforces a change in $\psi_5$ as well. This could be interpreted as a hint that D4$_1$ and D4$_5$ should effectively be regarded as a single stack of D-branes.

On top of that, these same two branes are also individually Freed--Witten anomalous, and become non-anomalous once there is a D2-brane suspended between them. Again, this would pair them into a new D4--D2--D4 system whose description is, at least to us, unknown. Similarly, one could pair D4$_2$ with D4$_6$ and D4$_3$ with D4$_7$, while the last remaining D4$_4$ would pair with the bulk D2-brane.

This coupling is also visible in the effective theory. The full classical untwisted scalar potential is
\begin{align}
V(\phi,s^i)
=
\frac{1}{16}
\Bigg[
&
\frac{m^2 e^{5\phi/2}}{\mathcal V^2}
+
\frac{h^2 e^{-\phi}}{\mathcal V^2}
\sum_{i=1}^{7}\frac{1}{(s^i)^2}
+
\frac{f^2 e^{\phi/2}}{\mathcal V^4}
\sum_{i=1}^{7}(s^i)^2
\nonumber\\
&
-
\frac{2mh\,e^{3\phi/4}}{\mathcal V^2}
\left(
\sum_{i=1}^{4}\frac{1}{s^i}
-
\sum_{i=5}^{7}\frac{1}{s^i}
\right)
+
\frac{2hf\,e^{-\phi/4}}{\mathcal V^3}
\Bigg],
\qquad
\mathcal V
=
\left(
\prod_{i=1}^{7}s^i
\right)^{1/3}.
\end{align}

It contains an explicit coupling between $f$ and $h$. Restoring $f$ therefore changes both the RR-flux energy and the source sector coupled to the residual $H_3$ background.

\begin{figure}[t]
\centering
\includegraphics[width=\linewidth]{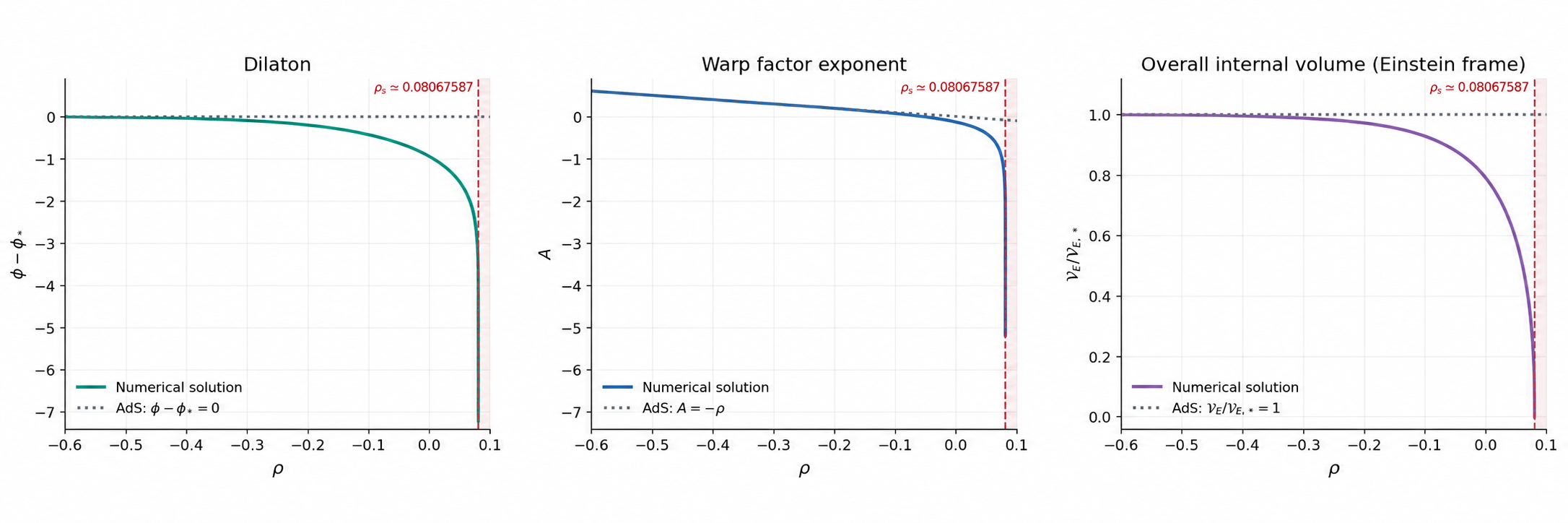}
\caption{Numerical BPS flow approaching the supersymmetric
$\mathrm{AdS}_3$ critical point. From left to right, the panels show the
dilaton, the warp-factor exponent, and the Einstein-frame internal volume
as functions of the radial variable $\rho$. The dotted curves indicate
the corresponding $\mathrm{AdS}_3$ reference behavior, while the dashed
red line marks the endpoint $\rho_s \simeq 0.08067587$ of the numerical
branch.}
\label{fig:ads3_flow}
\end{figure}

The flux-backtracking construction of
Refs.~\cite{Apers:2025pon, Apers:2026lgi} motivates placing
the D4 stacks on the background of the preceding subsection and attempting
to incorporate their backreaction through factorised harmonic
superposition. In the present model, the independent-stack version of this
prescription does not solve the full coupled flow equations: it misses the
correlated source response described above. It therefore does not yield a
full analytical interpolating metric. A flow approaching the supersymmetric $\mathrm{AdS}_3$
vacuum can nevertheless be obtained numerically by integrating the full
coupled BPS equations and selecting the branch that reaches the AdS critical
point, as shown in Figure~\ref{fig:ads3_flow}. 

\begin{table}[ht!]
\centering
\[
\begin{array}{c|c|c|c|c|c|c|c|c|c|c}
 & x_0 & x_1 & r & y_1 & y_2 & y_3 & y_4 & y_5 & y_6 & y_7 \\ \hline
 \text{D2} & \times & \times & \times & = & = & = & = & =  & = & =  \\ \hline 
\text{D4}_{1} & \times & \times & = & \times & = & \times & = & = & \times & = \\ \hline
\text{D4}_{2} & \times & \times & = & = & \times & = & \times & =  & \times & = \\ \hline
\text{D4}_{3} & \times & \times & = & = & \times & \times & = & \times  & = & = \\ \hline
\text{D4}_{4} & \times & \times & = & \times & \times & = & = & =  & = & \times \\ \hline
\text{D4}_{5} & \times & \times & = & = & = & = & = & \times  & \times & \times \\ \hline
\text{D4}_{6} & \times & \times & = & = & = & \times & \times & =  & = & \times \\ \hline
\text{D4}_{7} & \times & \times & = & \times & = & = & \times & \times  & = & =  
\end{array}
\]
\caption{D4-branes network with the D2-brane from the flux compactification. 
One can see that they are mutually supersymmetric, however the D4-branes are Freed-Witten anomalous on the $H_3$ background of the flux compactification. 
This means the brane picture here is not microscopically consistent and can only be used at best as a proxy for a fully fledged anomaly-free brane network.} 
\label{tab:branes}
\end{table}
As a final comment, although we do not have a microscopic description in terms of harmonic superposition, one can still ask what the naive brane counting would predict. Since $N_{\mathrm{D2}}$ grows with $f$, both the D4 and D2 sectors should be included; their worldvolume directions are summarised in Table~\ref{tab:branes}. Assigning a common parameter $n$ to the constituent D-brane contributions, and ignoring for the moment the correlations among them, the standard D-brane metric and dilaton weights give
\begin{equation}
    e^{\phi}_{\mathrm{count}}\sim n^{-3/2},
    \qquad
    \mathcal V_{S,\mathrm{count}}\sim n^{7/2},
    \qquad
    s^i_{\mathrm{count}}\sim n^{21/8},
\end{equation}
and hence
\begin{equation}
    c_{\mathrm{count}}
    \sim
    \frac{L_{\mathrm{AdS},S}}{\ell_s}
    \frac{\mathcal V_{S,\mathrm{count}}}
         {e^{2\phi}_{\mathrm{count}}}
    \sim n^8.
\end{equation}

These scalings do not match those of the actual flux vacuum. For $f=(2\pi)^3N$ with $m,h$ fixed, one instead finds
\begin{equation}
    e^{\phi}_{\mathrm{vac}}\sim N^{-3/4},
    \qquad
    \mathcal V_{S,\mathrm{vac}}\sim N^{7/4},
    \qquad
    s^i_{\mathrm{vac}}\sim N^{21/16},
    \qquad
    c_{\mathrm{vac}}\sim N^4.
\end{equation}
The two sets of exponents are related by the replacement
\begin{equation}
    n\longmapsto N^{1/2}.
\end{equation}
We do not know of a microscopic reason for this square-root rescaling, and it may simply be a coincidence of the parametric exponents. Given that the D4-branes are correlated by both tadpole cancellation and Freed--Witten anomaly cancellation, the naive independent-brane counting is in any case not expected to provide a faithful microscopic description.

\section{Conclusions}

In this work we have checked and verified the consistency of a series of detailed properties for G2 orientifold flux compactifications of massive IIA supergravity. 
This includes full moduli stabilization in the untwisted and twisted sectors (always in the supergravity regime of validity), 
curvature corrected tadpole cancellation, 
and elimination of orientifold intersections. 
This certainly puts the G2 orientifold constructions of IIA with scale separation on a firmer ground. 
We also performed a preliminary check for flux backtracking which showed that the microscopic D-brane picture is much more intricate than in the 4D cousins of these models. 
However, these models can have instead some interesting features such as finite interval solution of the resulting flow equations and the desirable low-energy decoupling of brane modes.

\section*{Acknowledgments}

We thank Thomas Van Riet for discussions. AA is supported by the Czech Science Foundation
GACR grant “On and Off-shell Superspace Representations” (GA26-23375S). F.A. is supported by the ERC Starting Grant QGuide101042568 -- StG 2021.

\appendix 

\section{Eguchi--Hanson space and its properties}
\label{Appendix Eguchi Hanson}

The Eguchi--Hanson space is a ALE space with a metric 
\begin{equation}
d s^2=\Delta^{-1} d r^2+\frac{1}{4} r^2 \Delta(d \psi+\cos \theta d \phi)^2+\frac{1}{4} r^2 d \Omega_2^2 \,, 
\end{equation}
with $\Delta=1-(\rho_0 / r)^4$ and $d \Omega_2^2=d \theta^2+\sin ^2 \theta d \phi^2$. 
The periodicity of $\psi =2 \pi$ instead of the usual $4 \pi$ in case of $\psi$ as a fiber angle of $S^3$ in Hopf fiber coordinates, 
making the $\mathbb{Z}_2$ symmetry manifest. 
The parameter $\rho_0$ denotes the radius of the Eguchi--Hanson exceptional sphere. 
At $r\rightarrow \rho_0$ we recover a smooth manifold, metric of $S^2$ with the radius $\rho_0$, this is called a bolt or the exceptional sphere.
Close to the bolt where we define 
$r= \rho_0+ \frac{\rho^2}{\rho_0} $ the metric looks like 
\begin{equation}
    ds^2_{EH}= d\rho^2 +\rho^2(d\psi +\cos\theta d\phi)^2+ \rho^2 (d\theta +\sin\theta d\phi)^2 \,, 
\end{equation}
which is locally a $\mathbb{R}^2\times S^2$ space. 
This real 4D space can also be written in terms of 2D complex coordinates by defining 
\begin{equation}
    z_1= x_1+ix_2 , \quad z_2= x_3+ix_4 \,. 
\end{equation}
In a more convenient coordinate system one can obtain the Eguchi--Hanson metric from the K\"ahler potential of form  
\begin{equation}
\mathcal{K}=\sqrt{u^4+\rho_0^4}+2 \rho_0^2 \ln u-\rho_0^2 \ln \left(\sqrt{u^4+\rho_0^4}+\rho_0^2\right) , 
\ \text{where} \ \ u^2=\left|z_1\right|^2+\left|z_2\right|^2 \,, 
\end{equation}
and we get
\begin{equation}
ds_{\mathrm{EH}}^2=\frac{\mathcal{F}^{\prime}}{2 u} d u^2+\frac{u \mathcal{F}^{\prime}}{2} \sigma_1^2+\mathcal{F}\left(\sigma_2^2+\sigma_3^2\right) \,. 
\end{equation}
Here the $\sigma_i$ are the standard Maurer--Cartan forms on $S^3$. 
The standard hyper-k\"ahler form on this space is given by $w_1$, associated with the K\"ahler potential, and reads 
\be 
w_1=\frac{\mathcal{F}^{\prime}}{2} d u \wedge \sigma_1+\mathcal{F} \sigma_2 \wedge \sigma_3 \,. 
\ee 
The two other K\"ahler forms, which must exist on a hyper-k\"ahler space, are given by 
\be 
w_2=\operatorname{Re}\left(d z_1 \wedge d z_2\right) \, , \quad w_3=\operatorname{Im}\left(d z_1 \wedge d z_2\right) \,. 
\ee 
Since transversal to the EH space (due to the embedding in the G2 orbifold) we have a 3-tori we can now define a 3-form and a dual 4-form. 
There is an anti-self dual closed 2-form that can exist on the bolt given by 
\begin{equation}\label{2 form}
\nu=\frac{\mathcal{F}^{\prime}}{2 \mathcal{F}^2} d u \wedge \sigma_1+\frac{1}{\mathcal{F}} \sigma_2 \wedge \sigma_3. 
\end{equation}
Near a $\Theta_\alpha$ fixed torus, we group the transverse and tangential directions as 
\be 
\left(x^1, x^2, x^3, x^4\right)=\left(y^1, y^2, y^3, y^4\right), \quad\left(\xi^1, \xi^2, \xi^3\right)=\left(y^5, y^6, y^7\right) \, , 
\ee
and the associative three-form then reads (locally) 
\begin{equation}\label{LocalPhi}
\Phi_{\mathrm{loc}}=\sum_{a=1}^3 \omega_a^{\mathrm{EH}} \wedge d\xi^a-d \xi^1 \wedge d \xi^2 \wedge d \xi^3 .
\end{equation}
Since each $\omega_a^{\mathrm{EH}}$ is closed we have $d \Phi_{\mathrm{loc}}=0$. 

In the Eguchi--Hanson core, the triplet is hyperkähler, and $\Phi_{\text {loc }}$ is torsion-free.
The untwisted associative form in the bulk is given by 
\be 
\Phi=-d y^{567}+d y^{127}+d y^{136}-d y^{347}-d y^{235}+d y^{145}+d y^{246} .
\ee 
which can be rewritten as
\be 
\label{Phi0}
\Phi_0=\omega_1^{\text {flat }} \wedge d \xi^1+\omega_2^{\text {flat }} \wedge d \xi^2+\omega_3^{\text {flat }} \wedge d \xi^3-d \xi^{123},
\ee 
where
\be
\begin{aligned}
& \omega_1^{\text {flat }}=d y^{14}-d y^{23}, \\
& \omega_2^{\text {flat }}=d y^{13}+d y^{24}, \\
& \omega_3^{\text {flat }}=d y^{12}-d y^{34} .
\end{aligned}
\ee 
These form a flat hyper-K\"ahler triplet on the transverse $\mathbb{R}^4$. 
The local resolved G2 form is obtained by replacing this flat triplet by the smoothed Eguchi--Hanson triplet. With a suitable choice of interpolating function modifying the K\"ahler potential one can show that three forms on localized Eguchi--Hanson maps to flat global three forms, this was done in \cite{Joyce2}. One also needs to choose an embedding of Eguchi Hansen space by choosing a particular orientation and show that these map to global G2 forms, this was done explicitly in \cite{Lukas:2003dn}. 
The volume of the resolved G2 manifold follows from the G2 structure forms by definition, with the standard normalization
\begin{equation}
\int_{X_7} \Phi \wedge \Psi=7 \mathcal{V} .
\end{equation}
where $\Psi \equiv \star_{\Phi} \Phi$, the hodge dual is taken with respect to the metric constructed from $\Phi$ itself, making this a non linear map.

What is clear is the fact that with this procedure we have 3 blow-up moduli corresponding to each EH resolution, one radius of blow-up, 2 orientation of the EH space in the manifold. The relevant moduli might be truncated to some subset of these blow up moduli which would depend on the orientifold planes, we make this distinction clear in~\ref{Counting the active ingredients}.

\section{More examples and flux choices}

\subsection{Example without D6-branes and D2 scaling}\label{Solution1}

We give another example and work out some of the main details. 

\subsubsection{Fluxes} 

Now we are ready to insert fluxes and pick signs such that tadpole cancellation happens consistently. 
The Romans mass is 
\be
F_0 = m > 0 \ , 
\ee
and the $F_4$ flux can be taken fully isotropic because of the scaling D2 
\be \label{F_4 flux choice}
F_4 = - f \sum_i \psi_i  \quad , \quad f > 0 \ . 
\ee
The $H_3$ is essentially non-isotropic because it cancels only O6 tadpoles 
\be
\int_i H_3 \wedge F_0 = - (2 \pi) (N_{\rm D6,i} -2 N_{\rm O6,i})\,,  
\ee
where the $-2$ inside the parenthesis corresponds to the negative charge of an O6 plane and therefore for $N_{\rm D6,i}=0$ we have 
\be \label{H_3 flux choice}
H_3 = h (\phi_1+\phi_2+\phi_3+\phi_4)  \quad , \quad h > 0 \ .  
\ee
The D2 tadpole further gives 
\be
N_{\rm D2} = \frac{4 hf}{(2 \pi)^5} + 2 \,. 
\ee
We see that at large $f$ the number of D2-branes also increases with $f$.

Inserting these ingredients in the superpotential gives 
\be \label{Superpotential bulk}
P = \frac{m e^{5\phi/4}}{8 {\cal V}} 
+ \frac{h e^{-\phi/2}}{8 {\cal V}} \left( \frac{1}{s_1} + \frac{1}{s_2} + \frac{1}{s_3} + \frac{1}{s_4} \right)  
- \frac{f e^{\phi/4}}{8 {\cal V}^2} \sum_{i=1}^7 s_i 
\,, 
\ee 
with ${\cal V} = (s_1 s_2 s_3 s_4 s_5 s_6 s_7)^{1/3}$. 
Once we extremize the superpotential with respect to the eigth moduli we find that the dilaton is 
\be
e^\phi = \frac{2^{1/4} 4 }{3} \frac{h}{m^{1/4} f^{3/4}} \,, 
\ee
and the shape moduli are 
\be
s_{1,2,3,4} = \frac{8h}{3m} e^{-7 \phi/4} \quad , \quad s_{5,6,7} = \frac{32h}{3m} e^{-7 \phi/4} \,. 
\ee
This guarantees parametric large volume (both in Einstein and string frame) and parametrically weak coupling at large $f$. 
The analysis for scale separation carries through exactly as in \cite{Farakos:2020phe}. Therefore at, large $f$, the AdS$_3$ vacuum is scale separated 
\be
\frac{L_{\rm KK}^2}{L_{\rm AdS}^2} \sim f^{-1} \,. 
\ee
Note that the string frame volume scales like $f^{7/4}$ and since the number of D2-branes increases with $f$ we have 
\be
\frac{N_{\rm D2}}{{\cal V}_S} \sim f^{-3/4} \,, 
\ee
therefore they are diluted in the scale separation limit. 

We can also verify full moduli stabilization by the fact that all masses are non-zero, 
and we can also evaluate the conformal dimensions of the dual operators for the scalars which read 
\be
\Delta_i = \left\{1,1,1,4,4,4, \frac{\sqrt{73} - 3}{2}, \frac{\sqrt{73} + 7}{2} \right\} \,. 
\ee
It is intriguing to see that six out of the eight are integers. 
There may be some deformation that leads to exact integer dimensions but we have not found it within the few cases we tried.

\subsubsection{Blow-up stabilization}

Now we work out the blow-up moduli stabilization for the model in \ref{Solution1} explicitly and show how an appropriate choice of exceptional fluxes scaling can lead to parametrically small blow-up radius compared to bulk moduli. We have already emphasized that apart from the numerics the scaling and arguments hold for any model but it is still nice to see it working in the simplest model which is \ref{Solution1}, solidifying our trust in the approach. WLOG, we will only consider the blow-up of fixed points of $\Theta_{\alpha}$ involution. Corresponding physical moduli comes from the $u_1,u_3$ cycles as discussed in sec.~\ref{Counting the active ingredients} and \ref{A specific example}

For the flux choice as defined in eq.\eqref{F_4 flux choice} and eq.\eqref{H_3 flux choice}-
\begin{equation}
f^i=-f\,(1,1,1,1,1,1,1),
\qquad
h^i=h\,(1,1,1,1,0,0,0),
\qquad
F_0=m,
\end{equation}
with $f,h,m>0$. 
The resolved Einstein-frame volume, to quadratic order in the exceptional
periods, is
\begin{equation}
\mathcal V 
= V_0\left(1-\frac{8}{3}\sum_{a=1}^3d_au_a^2\right)
+{\cal O}(u^4),
\qquad
V_0=\left(\prod_{i=1}^7s^i\right)^{1/3},
\end{equation}
where 
\begin{equation}
d_a=\frac{1}{s^{A_a}s^{B_a}},
\qquad
(A_a, B_a)=(5,7),(1,6).
\end{equation}

Expanding eq \eqref{Superpotential correction} around the orbifold locus takes the form for the given flux choice we get 
\begin{equation}
P
=
P_{0}
+
\sum_{a=1}^3L_au_a
+
\frac{1}{2}\sum_{a=1}^2X_au_a^2
+
{\cal O}(u^3).
\end{equation}
Here $P_0$ is the superpotential of the bulk fields given in eq.\eqref{Superpotential bulk}.
The exceptional-flux coefficient linear in order is  
\begin{equation}
L_a
=
\frac{1}{8}e^{\phi/4}V_0^{-2}q_a,
\end{equation}
while the quadratic coefficient is
\begin{equation}
X_a
=
d_a\Bigg\{
\frac{2}{3}m e^{5\phi/4}V_0^{-1}
+
e^{-\phi/2}V_0^{-1}
\left[
\frac{2}{3}\sum_{i=1}^7\frac{h^i}{s^i}
+
\frac{h^{A_a}}{s^{A_a}}
+
\frac{h^{B_a}}{s^{B_a}}
\right]
+
\frac{4}{3}e^{\phi/4}V_0^{-2}
\sum_{i=1}^7f^is^i
\Bigg\}.
\label{eq:Ba-general-new-flux}
\end{equation}

For the untwisted supersymmetric solution is given in section\eqref{Solution1}. We plug in the scalings of the bulk moduli assuming the correction to the superpotential does not affect the leading order solution. 
The corresponding Einstein-frame volume is
\begin{equation}
V_0
=
3^{7/4}\times2^{-3/16}
\frac{f^{49/16}}{m^{21/16}h^{7/4}}.
\end{equation}
Substitution into \eqref{eq:Ba-general-new-flux} gives
\begin{equation}
B_1=B_3=
-\frac{2^{7/8}}{3^{9/2}}
\frac{h^{9/2}m^{25/8}}{f^{53/8}}.
\end{equation}
In particular, all three exceptional periods receive nonvanishing and
equal quadratic coefficients.

The leading supersymmetry equations in the exceptional sector are
\begin{equation}
P_{u_a}
=
L_a+X_au_a+O(u^2)=0 \,, 
\end{equation}
while give the value for the blow-up moduli 
\begin{equation}
u_a^\star=-\frac{L_a}{X_a}.
\end{equation}
Equivalently, in terms of the bulk periods entering the corresponding
volume correction,
\begin{equation}
\frac{u_1^\star}
{\sqrt{s^{A_1}s^{B_1}}}
=
\frac{q_a}{8\,f} \,, 
\label{eq:ua-new-flux}
\end{equation}
and similarly 
\begin{equation}
     \frac{u_3^\star}
{\sqrt{s^{A_3}s^{B_3}}}
=
\frac{q_a}{4\,f} \,.
\end{equation}
The relative correction to the Einstein-frame volume is therefore
\begin{equation}
\frac{\Delta V}{V_0}
=-\frac{1}{4f^2}\Big(
 \frac{q_1^2}{96} + \frac{q_3^2}{24}\Big)+
{\cal O}\left(\frac{Q_E^4}{f^4}\right) \,. 
\end{equation}
Thus the small-resolution expansion is controlled for
\begin{equation}
Q_E\ll f.
\end{equation}
We also define for convenience 
\begin{equation}
Q_E^2\equiv 16 q_1^2 +16 q_3^2,
\end{equation}
for a normalized flux quanta on the exceptional cycles.
We emphasize again that the untwisted solution is not an exact solution after turning on
$q_a\neq0$. Integrating out the blow-up modulus at leading order gives
\begin{equation}
\Delta P_\star
=
-\sum_{a=1}^3\frac{L_a^2}{2 X_a},
\end{equation}
and hence the shifts of the dilaton and untwisted geometric moduli begin
parametrically at
\begin{equation}
\frac{\delta s^i}{s^i} \sim  \delta\phi \sim  {\cal O} \left(\frac{Q_E^2}{f^2}\right) \,. 
\end{equation}
While stabilization of the blow-up cycle at a parametrically smaller scale than the bulk moduli is desirable feature, one carefully analyze if the size of the blow-up cycles remains larger then the string length so that the supergravity approximation is still valid.

Here our analysis follows the same conclusion as one in sec.~\ref{Stabilization of blow-up modes} and \ref{A specific example},
The characteristic bulk radius scales as
\begin{equation}
\frac{R_S}{\ell_s}
\sim
\left(\frac{f}{m}\right)^{1/4}.
\end{equation}
Using the standard Eguchi--Hanson period relation in eq.\eqref{Periods relation}, the exceptional cycle scale behaves as 
\begin{equation}
\frac{\rho_S^2}{\alpha'}
\sim
\frac{Q_{\mathrm{E}}}{\sqrt{fm}},
\end{equation}
For a fixed $m$, parametrically controlled resolved regime therefore requires
\begin{equation}
\sqrt{f}\ll Q_{\mathrm{E}}\ll f.
\end{equation}
This regime can be chosen for appropriate scaling of $Q_E$ with the flux quant $f$.
We hence state that the above model also remains scale separated at parametrically large flux choice on a smooth non singular G2 manifold.

\subsubsection{Flux backtracking}

We now apply the flux-backtracking procedure to this model.
To determine the background probed by the D4-branes sourcing the $F_4$ fluxes, we remove
these fluxes from the superpotential while retaining the Romans mass and the NSNS flux,
\begin{equation}
F_0=m,\qquad H_3=h(\phi_1+\phi_2+\phi_3+\phi_4), \qquad F_4=0.
\end{equation}
The resulting flow is the one studied in the main text with $H_-=0$: only the Romans
mass and $H_+$ contributions remain. Hence, we find the ten-dimensional string frame metric,
\begin{align}
ds_{10,S}^2={}&d\rho^2+\rho^{-10/9}(-dt^2+dx^2)\notag\\
&+\rho^{2/3}\bigl[(dy^1)^2+(dy^3)^2+(dy^6)^2\bigr]
+\rho^{14/9}(dy^2)^2\notag\\
&+\rho^{-2/9}\bigl[(dy^4)^2+(dy^5)^2+(dy^7)^2\bigr],
\qquad e^\phi\propto\rho^{-1}.
\end{align}
This is precisely the scaling solution found near the left endpoint in
Section 5.1.3, where $H_-$ is subleading. Since $H_-$ is absent here, that
solution is exact globally on $0<\rho<\infty$, rather than merely asymptotic.
The flow therefore extends to arbitrarily large radius, with no second endpoint
or finite radial cutoff.

\subsection{Flux example with D6-branes and no D2 scaling}

We can also work-out a setup where the D2 number does not scale and so $F_4 \wedge H_3$ = 0. 
There are various way to achieve such a condition and we have not exhausted the possibilities however we will report on one specific working example just to make the discussion clear. 
The Romans mass is 
\be
F_0 = m > 0 \ , 
\ee
and the $F_4$ flux is not fully isotropic (one could also have isotropic flux by altering a bit the D6-branes) 
\be
F_4 = - 3 f (\psi_1 + \psi_2 + \psi_3 + \psi_4) - 4 f (\psi_5 + \psi_6 + \psi_7) \quad , \quad f > 0 \ , 
\ee
and the $H_3$ is essentially non-isotropic because of O6/D6 planes 
\be
H_3 = h (\phi_1 + \phi_2 + \phi_3 + \phi_4) - h (\phi_5 + \phi_6 + \phi_7) \quad , \quad h > 0   \ .  
\ee
We see that the $H_3$ part related to the O6 tadpole cancellation remains the same whereas the new part has opposite sign and is related to the D6 tadpole. 
Since D6-branes carry half the charge of O6-planes we can certainly have equal number of total D6-brane charge per cycle, in the relevant cycles, as we have for O6-planes in the relevant cycles. 
Note also that the D2 tadpole now just requires a specific fixed number of D2-branes 
\be
N_{\rm D2} = 2 \,. 
\ee
The extremization of the superpotential does not lead to any surprises and we still get the same scalings for the dilaton and the shape moduli (only some numerical coefficients change) while we can be at parametric large volume and weak coupling at the scale separation limit. 
The model provides full moduli stabilization but the dual dimensions are nowhere near integers so we do not  record them here explicitly. 
It goes without saying that one could maybe find some deformation that provides integers but we have not found one for this setup with the few cases we tried.

\end{document}